\documentclass[aps,twocolumn]{revtex4}

\RequirePackage[T2A]{fontenc}

\usepackage{amssymb}
\usepackage{amsmath}
\usepackage{bm}
\usepackage[dvips]{epsfig}
\usepackage{graphicx}

\begin{document}

\title{On the level lines of two-layer symmetric potentials}

\author{A.Ya. Maltsev}

\affiliation{
\centerline{\it L.D. Landau Institute for Theoretical Physics}
\centerline{\it 142432 Chernogolovka, pr. Ak. Semenova 1A,
maltsev@itp.ac.ru}}

\begin{abstract}
 We consider the behavior of level lines of two-dimensional 
potentials, which play an important role in the physics 
of ``two-layer'' systems. Potentials of this type are 
quasiperiodic and, at the same time, can also be considered 
as a model of random potentials on a plane. The description 
of level lines of such potentials is a special case of 
the Novikov problem for potentials with four quasiperiods 
and uses many features that arise in the study of the 
general Novikov problem. At the same time, the potentials 
under consideration also have their own clearly expressed 
specificity, which makes them very interesting for research 
from a variety of points of view.
\end{abstract}

\maketitle

\section{Introduction}

 The problem considered here is a special case of the 
problem of S.P. Novikov, namely, the problem of describing 
the level lines of quasiperiodic functions on a plane. 
Note immediately that a quasiperiodic function on a 
plane $\, \mathbb{R}^{2} \, $ with $\, N \, $ 
quasiperiods will mean here the restriction  
of an $\, N \, $-periodic function 
$\, F (z^{1}, \dots , z^{N}) \, $ in the space 
$\, \mathbb{R}^{N} \, $ to the plane 
$\, \mathbb{R}^{2} \, $ under a generic affine embedding 
$\, \Pi \, = \, \mathbb{R}^{2} \, \subset \, \mathbb{R}^{N} \, $.

 The Novikov problem was first set in 
\cite{MultValAnMorseTheory} for the case $\, N = 3 \, $, 
and it is for this case that it has been studied most 
deeply (see, for example, 
\cite{zorich1,dynn1992,Tsarev,dynn1,zorich2,DynnBuDA,dynn2,dynn3}). 
In addition to the theory of two-dimensional systems, 
the case $\, N = 3 \, $ also plays, in fact, an important role 
in the theory of galvanomagnetic phenomena and is responsible 
for the emergence of nontrivial topological quantities and 
regimes observed in the magnetoconductivity of normal metals 
(see, for example, 
\cite{PismaZhETF,ZhETF2,UFN,BullBrazMathSoc,JournStatPhys,TrMian}).

 The problem we consider here corresponds to the case 
$\, N = 4 \, $. The deepest general results for this case 
were obtained in the works \cite{NovKvazFunc,DynNov}. 
It should be said, however, that in general 
the case $\, N = 4 \, $ has not been studied 
as deeply as the case $\, N = 3 \, $.

 The main part of Novikov's problem (as well as the basis 
for its consideration) is the problem of describing open 
(non-closed) level lines of the function
$$f (x, y) \,\,\, = \,\,\, 
F ({\bf z}) \big|_{\Pi = \mathbb{R}^{2} \subset  \mathbb{R}^{N}} $$

\begin{figure}[t]
\begin{center}
\includegraphics[width=\linewidth]{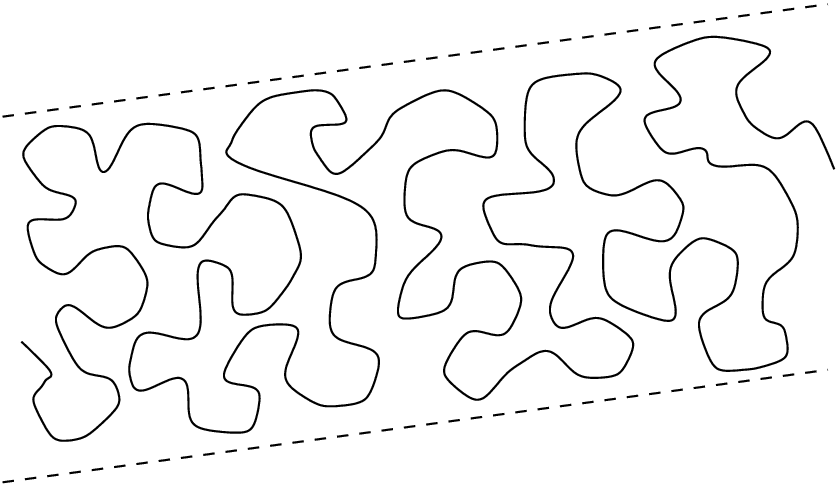}
\end{center}
\caption{The form of a ``topologically regular'' open 
level line of a quasi-periodic function $\, f (x, y) \, $}
\label{TopReg}
\end{figure}

 An important role is played by the division 
of open level lines
\begin{equation}
\label{cLevels}
f (x, y) \,\,\, = \,\,\, c
\end{equation}
into ``topologically regular'' and ``chaotic''. 
The difference between ``topologically regular'' 
and ``chaotic'' open level lines (\ref{cLevels}) 
is manifested primarily in their global 
behavior on large scales. Namely, 
each ``topologically regular'' open level 
line (\ref{cLevels}) lies in a straight strip 
of finite width in the plane $\, \Pi \, $, 
passing through it (Fig. \ref{TopReg}) 
(note that ``topologically regular'' level 
lines, generally speaking, are not periodic).

\begin{figure}[t]
\begin{center}
\includegraphics[width=\linewidth]{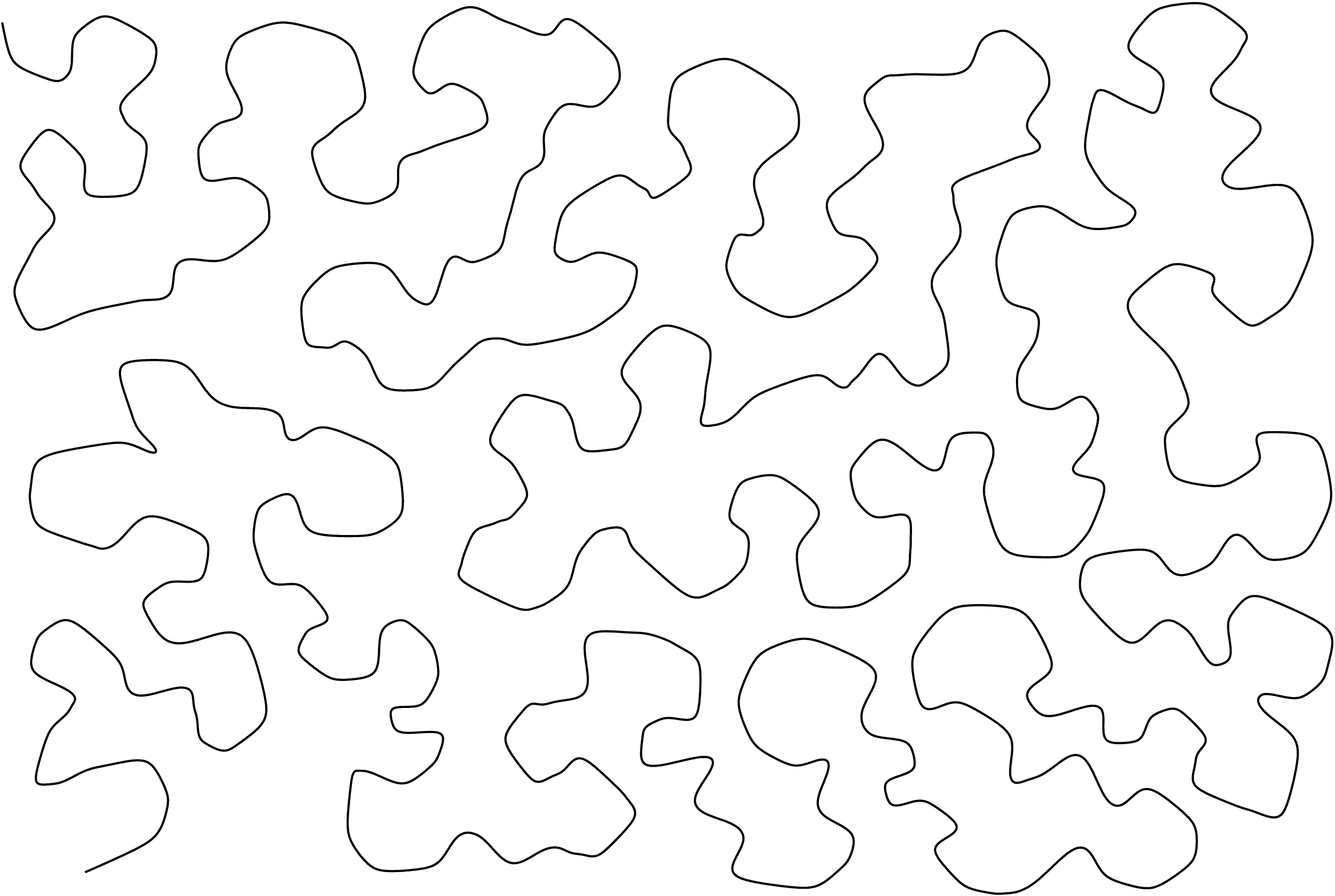}
\end{center}
\caption{The form of a ``chaotic'' open level line of 
a quasiperiodic function $\, f (x, y) \, $ (schematically)}
\label{Chaotic}
\end{figure}

 ``Chaotic'' open level lines exhibit more complex 
behavior, wandering ``everywhere'' in the 
plane $\, \Pi \, $ (Fig. \ref{Chaotic}).

 Another difference between ``topologically regular'' 
open level lines (\ref{cLevels}) and ``chaotic'' ones 
is their stability with respect to small variations 
of the problem parameters. In particular, 
``topologically regular'' open level lines (\ref{cLevels}), 
as a rule, are preserved (and preserve their mean
direction) with small variations of the parameter 
$\, c \, $, as well as the function $\, F ({\bf z}) \, $. 
In addition, ``topologically regular'' level lines 
(\ref{cLevels}), arising in one of the planes 
$\, \Pi \subset \mathbb{R}^{N} \, $, also arise in all 
other planes $\, \Pi^{\prime} \subset \mathbb{R}^{N} \, $ 
of the given direction.

  Despite such strong requirements, 
``topologically regular'' level lines (\ref{cLevels}) 
arise in many families of quasiperiodic functions, 
and in many cases represent the main type of open level 
lines (\ref{cLevels}), which is due to the general 
features of quasiperiodic functions on the plane.

\vspace{1mm}

 The ``chaotic'' level lines, on the contrary, are 
usually unstable to small variations of the problem 
parameters. For example, for $\, N = 3\, $ ``chaotic'' 
lines (\ref{cLevels}) can arise only at a single level 
$\, c \, $ (see \cite{dynn1}), while the level lines 
of $\, f (x, y) \, $  at all other levels are closed. 
``Chaotic'' level lines have complex ``scaling''
properties (see \cite{DynnBuDA,dynn2,Zorich1996,
ZhETF2,ZorichAMS1997,Zorich1997,zorich3,ZorichLesHouches,
Skripchenko2,DynnSkrip1,DynnSkrip2,AvilaHubSkrip1,TrMian}), 
which makes the corresponding functions $\, f (x, y) \, $ 
related to various models of random potentials on the plane. 
(In this sense, the functions $\, f (x, y) \, $, 
having ``topologically regular'' open level lines, 
are more similar to periodic potentials).

\vspace{1mm}

 In most cases, the emergence of ``topologically regular'' 
level lines (\ref{cLevels}) corresponds to a union of 
(a finite or infinite number) some open sets 
(Stability Zones) in the parameter space of the problem, 
while the emergence of ``chaotic'' level lines 
(\ref{cLevels}) corresponds to Cantor-type sets in 
the parameter space (see, for example, 
\cite{zorich1,dynn3,NovKvazFunc,DynNov,DeLeo1,DeLeo2,
DeLeo3,DeLeoDynnikov1,dynn4,DeLeoDynnikov2,Skripchenko1,
AvilaHubSkrip2,AnnPhys,DynHubSkrip}).

\vspace{1mm}

 The potentials $\, V (x, y) \, $ considered here 
represent a superposition of periodic potentials 
$\, V_{1} (x, y) \, $ and $\, V_{2} (x, y) \, $, where 
\begin{equation}
\label{V2Pot}
V_{2} ({\bf r}) \,\,\, = \,\,\, \overline{V}_{1} 
\Big( \pi_{-\alpha} ({\bf r} - {\bf a}) \Big) 
\end{equation}
represents the rotation and the shift of the potential 
$\, V_{1} (x, y) \, $ reflected about the $\, x$ - axis 
by some angle $\alpha \, $ and vector $\, {\bf a} \, $, 
respectively (here we also denote by 
$\, \pi_{\alpha} \left({\cal F}\right) \, $ the rotation 
of any figure $\, {\cal F} \, $ in $\, \mathbb{R}^{2} \, $ 
by an angle $\alpha \, $ relative to the origin).

 The above potentials belong to the case $\, N = 4 \, $ 
and can be considered from the point of view of the general 
Novikov problem (see, e.g., \cite{AnnPhys}). Here, however, 
we impose additional conditions on the potential 
$\, V_{1} (x, y) \, $, which gives additional specificity 
to this problem. Namely, we require that the potential 
$\, V_{1} (x, y) \, $ has rotational symmetry (of the third, 
fourth, or sixth order) in the plane $\, \mathbb{R}^{2} \, $. 
(Without loss of generality, we will consider here the origin 
to be one of the centers of rotational symmetry 
of $\, V_{1} (x, y) $).

 Potentials of this type arise in ``two-layer'' systems, 
when a monatomic layer is superimposed on an ``inverted'' 
identical layer with some rotation and shift in 
$\, \mathbb{R}^{2} \, $. An important subclass of such 
potentials is given by systems where the potential 
$\, V_{1} (x, y) \, $ has reflection symmetry 
(with respect to some axis) and the set of potentials 
$\, V_{2} (x, y) \, $ coincides, in fact, with the set
$$V_{1} \Big( \pi_{-\alpha} ({\bf r} - {\bf a}) \Big) $$

 As is well known, potentials of this type are typical 
for the physics of ``bilayer'' systems, of which bilayer 
graphene is one of the most important (we cite here only 
a few of the huge number of works on this topic 
\cite{Shallcross1,Shallcross2,GeimGrigorieva,RShRN,CMFCLK,
KhalafKruchTarnVish,DindorkarKuradeShaikh,PaulCrowleyFu,
BernevigEfetov}). We also note that in a number of 
problems arising in such systems, the level lines 
of such potentials play an important role 
(see, e.g. \cite{TitovKatsnelson}).

\vspace{1mm}

 For special (``magic'') rotation angles $\, \alpha \, $ 
the potentials $\, V (x, y) \, $ are periodic. For generic 
angles $\, \alpha \, $ the potential $\, V (x, y) \, $ 
is a quasiperiodic function with 4 quasiperiods. In 
the latter case the potentials $\, V (x, y) \, $,
as can be shown, cannot have ``topologically regular'' 
level lines and their open level lines are of ``chaotic'' 
type. As follows from the results of \cite{Superpos}, 
open level lines of such potentials can arise only for 
a single energy value ($c = c_{0}$), which also brings 
such potentials closer to ``random'' potentials on the 
plane.

 In this paper we are interested in the behavior of 
(closed) level lines (\ref{cLevels}) at energy levels 
close to the level of emergence of open level lines. 
This problem is well known in the theory of random 
potentials, as well as in percolation theory, and, 
in particular, can play an important role when 
considering ``two-layer'' systems of finite sizes. 
The results obtained here allow, in particular, 
to reveal some specific features of the potentials 
under consideration in comparison with the general 
theory of ``random'' potentials on a plane.

\vspace{1mm}

 The potentials $\, V (x, y, \alpha, {\bf a}) \, $ 
studied here are a special case of the more general 
``two-layer'' potentials considered in \cite{Superpos} 
and corresponding to superpositions of potentials 
$\, V_{1} (x, y) \, $ and $\, V_{2} (x, y) \, $ of 
the same rotational symmetry. Here we impose 
the condition (\ref{V2Pot}) as an additional constraint.

 The rotational symmetry of the potential 
$\, V_{1} (x, y) \, $ can be of the third, 
fourth or sixth order. Here we consider in detail 
the case of the third-order symmetry, including 
the sixth-order symmetry as a special case. 
The case of the fourth-order symmetry is considered 
without any significant changes.

\section{General facts about level lines of quasiperiodic 
functions and potentials $\, V (x, y, \alpha, {\bf a}) \, $ }
\setcounter{equation}{0}

 As we have already said, we will consider here 
superpositions of the potential $\, V_{1} (x, y) \, $ 
and the potential $\, V_{2} (x, y) \, $, given by the 
formula (\ref{V2Pot}). For simplicity, we will consider 
here ``local'' superpositions of the form
$${\rm (1)} \quad \quad
V (x, y) \,\,\, = \,\,\, V_{1} (x, y) 
\,\, + \,\, V_{2} (x, y) $$
(simple linear superposition) or
$${\rm (2)} \quad V (x, y) \,\,\, = \,\,\, Q \, \Big( 
V_{1} (x, y), \, V_{2} (x, y) \Big) $$
for some smooth symmetric function
$\, Q \left( V_{1}, V_{2} \right) \, = \, 
Q \left( V_{2}, V_{1} \right) \, $
(nonlinear local superposition).

 The quasi-periodic structure of the potential 
$\, V (x, y) \, $ is generally defined by the embedding
\begin{equation}
\label{Embedding}
\mathbb{R}^{2} \, \rightarrow \mathbb{R}^{4} \, : \quad
{\bf r} \,\,\, \rightarrow \,\,\, 
\left( 
\begin{array}{c}
{\bf r}  \\
\pi_{-\alpha} ({\bf r} - {\bf a})
\end{array} \right)  
\end{equation}

 As is easy to see, in cases (1) and (2) the function 
$\, F ({\bf z}) \, $ is given by the formulas
$$F (z^{1}, z^{2}, z^{3}, z^{4}) \,\,\, = \,\,\, 
V_{1} (z^{1}, z^{2}) \,\, + \,\, 
\overline{V}_{1} (z^{3}, z^{4}) $$
and
$$F (z^{1}, z^{2}, z^{3}, z^{4}) \,\,\, = \,\,\,  
Q \, \Big( V_{1} (z^{1}, z^{2}), \, 
\overline{V}_{1} (z^{3}, z^{4}) \Big) $$
respectively.

 In both cases, the function $\, F ({\bf z}) \, $ 
is a smooth 4 - periodic function of $\, {\bf z} \, $
with periods determined by the periods of the potential 
$\, V_{1} (x, y) \, $. In particular, we will assume 
here the relation 
\begin{equation}
\label{C1Otsen}
\left| \nabla_{\bf z} F ({\bf z}) \right| 
\,\,\, \leq \,\,\, C_{1} 
\end{equation}
for some constant $\, C_{1} \, $ for all values
of $\, {\bf z} \, $.

 As is easy to see, different values of $\, {\bf a} \, $ 
correspond to parallel shifts of the plane 
$\, \Pi = \mathbb{R}^{2} \, $ in the space 
$\, \mathbb{R}^{4} \, $ with a fixed embedding direction 
$\, \mathbb{R}^{2} \subset \mathbb{R}^{4} \, $. For many 
reasons, it is usually convenient to consider the Novikov 
problem at once for the entire family of planes 
$\, \mathbb{R}^{2} \subset \mathbb{R}^{4} \, $ of a given 
direction for a fixed function $\, F ({\bf z}) \, $. 
Thus, we will often consider here the behavior of the level 
lines at once for all potentials $\, V (x, y) \, $, 
differing only in the shift vector $\, {\bf a} \, $ 
for a fixed $\, \alpha \, $.

 In the most general case 
(see \cite{dynn3,DynMalNovUMN,BigQuas}), for the complete 
family of functions $\, f (x, y) \, $, corresponding to 
the complete set of planes $\, \Pi \subset \mathbb{R}^{N} \, $ 
of a given generic direction $\, \xi \in G_{N,2} \, $, 
the following statements are true:

\vspace{2mm}

\noindent
1) The set of values $\, c \, $, corresponding to the 
emergence of open level lines
$$f (x, y) \,\,\, = \,\,\, F ({\bf z}) \big|_{\Pi} 
\,\,\, = \,\,\, c $$
in at least one plane $\, \Pi \, $ of the direction 
$\, \xi \, $, forms a closed segment
$$c \,\,\, \in \,\,\, \left[ 
c_{1} (\xi) , \, c_{2} (\xi) \right] \,\,\, , $$ 
which can be reduced to one point
$\, c_{0} (\xi) = c_{1} (\xi) = c_{2} (\xi) \, $. 

\vspace{2mm}

\noindent
2) In the case $\, c_{2} (\xi) > c_{1} (\xi) \, $ 
open level lines of functions $\, f (x, y) \, $ arise 
(simultaneously) in all planes $\, \Pi \, $ of 
direction $\, \xi \, $ if
$$c \,\,\, \in \,\,\, \left( 
c_{1} (\xi) , \, c_{2} (\xi) \right) $$

\vspace{2mm}

\noindent
3) For $\, c = c_{1} (\xi) \, $, $\, c = c_{2} (\xi) \, $ 
or $\, c = c_{0} (\xi) \, $ open level lines of functions 
$\, f (x, y) \, $ may arise only for a part of planes 
$\, \Pi \, $ of direction $\, \xi \, $. In this case, 
however, functions $\, f (x, y) \, $ have closed level 
lines of arbitrarily large sizes in all $\, \Pi \, $ 
(for the corresponding values of $\, c $).

\vspace{2mm}

 In particular, a quasiperiodic function may not have 
open level lines for any value of $\, c \, $. In this case, 
however, it must have arbitrarily large closed level lines 
at some single level $\, c = c_{0} \, $.

\vspace{2mm}

 For 
$\, c \, \notin \, \left[ c_{1} (\xi) , 
\, c_{2} (\xi) \right] \, $ all functions $\, f (x, y) \, $ 
have only closed level lines $\, f (x, y) = c \, $ in all 
planes of direction $\, \xi \, $. The sizes of all such level 
lines in all $\, \Pi \, $ are limited by one constant 
$\, G (c) \, $ which, however, can tend to infinity for 
$\, c \rightarrow c_{1} (\xi) \, $ or 
$\, c \rightarrow c_{2} (\xi) \, $ 
(\cite{dynn3,DynMalNovUMN}).

\vspace{2mm}

 In the case of a superposition of two periodic potentials 
$\, V_{1} (x, y) \, $ and $\, V_{2} (x, y) \, $ the direction 
of embedding $\, \mathbb{R}^{2} \subset \mathbb{R}^{4} \, $ 
is determined by the rotation angle $\, \alpha \, $, so that
$$c_{1} \, = \, c_{1} (\alpha) \,\,\, , \quad 
c_{2} \, = \, c_{2} (\alpha) \,\,\, , \quad
c_{0} \, = \, c_{0} (\alpha) $$

\vspace{1mm}

 For values $\, c < c_{1} (\alpha) \, $ all planes
$\, \Pi \, $ contain a single unbounded (connected) 
component of the set
$$f (x, y) \,\,\, > \,\,\, c $$

\begin{figure}[t]
\begin{center}
\includegraphics[width=\linewidth]{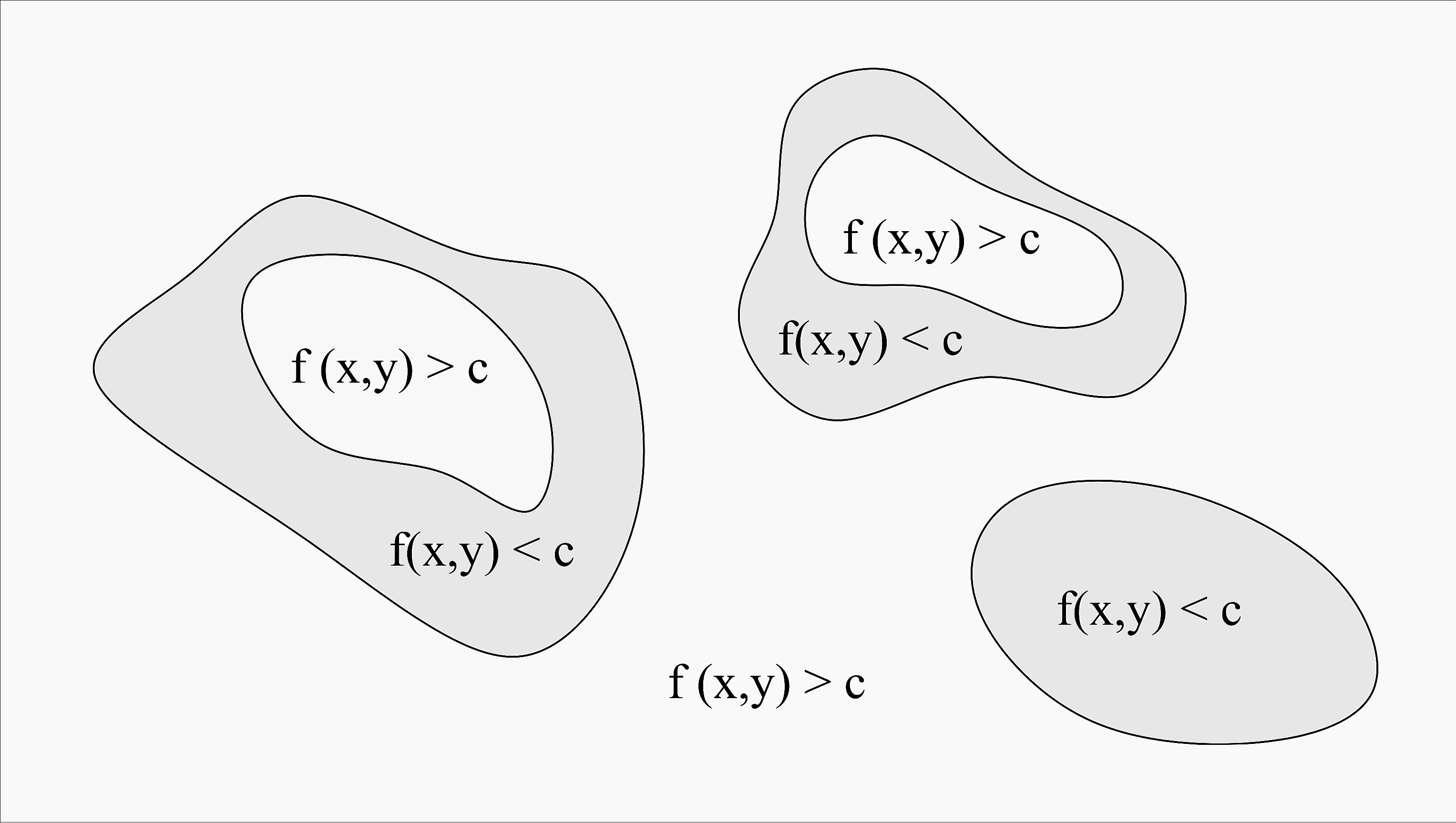}
\end{center}
\caption{Situation of type $\, A_{-} \, $ 
($c < c_{1} (\alpha)$) in the plane 
$\, \mathbb{R}^{2} \, $}
\label{SitAminus}
\end{figure}

 All other connected components of the sets 
$\, f (x, y) > c \, $ and $\, f (x, y) < c \, $ 
are bounded, and their sizes are limited by one 
constant $\, G (c) \, $ (Fig. \ref{SitAminus}).

 We will call such a situation in the plane $\, \Pi \, $ 
a situation of type $\, A_{-} \, $.

 Conversely, the presence of a situation of type 
$\, A_{-} \, $ in at least one plane $\, \Pi \, $ 
(at least for one value of $\, {\bf a}$) implies 
the relation $\, c < c_{1} (\alpha) \, $ for generic 
(not ``magic'') angles $\, \alpha \, $. 
 
\vspace{1mm}

 For values $\, c > c_{2} (\alpha) \, $ all planes
$\, \Pi \, $ contain a single unbounded (connected) 
component of the set
$$f (x, y) \,\,\, < \,\,\, c $$

\begin{figure}[t]
\begin{center}
\includegraphics[width=\linewidth]{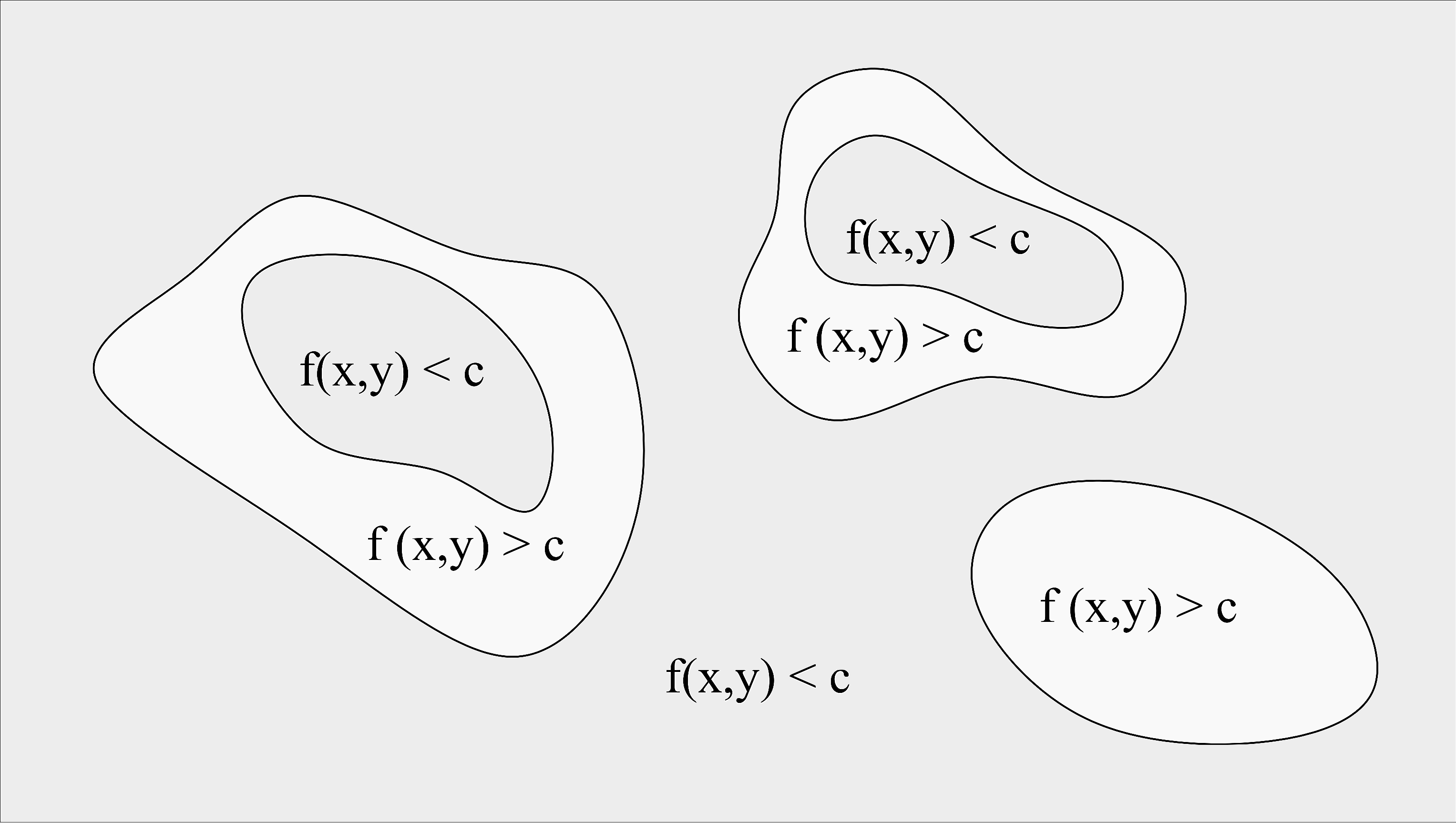}
\end{center}
\caption{Situation of type $\, A_{+} \, $ 
($c > c_{2} (\alpha)$) in the plane 
$\, \mathbb{R}^{2} \, $}
\label{SitAplus}
\end{figure}

 All other connected components of the sets 
$\, f (x, y) > c \, $ and $\, f (x, y) < c \, $ 
are bounded, and their sizes are limited by one 
constant $\, G (c) \, $ (Fig. \ref{SitAplus}).

 We will call such a situation in the plane $\, \Pi \, $ 
a situation of type $\, A_{+} \, $.

 Conversely, the presence of a situation of type 
$\, A_{+} \, $ in at least one plane $\, \Pi \, $ 
(at least for one value of $\, {\bf a}$) implies 
the relation $\, c > c_{2} (\alpha) \, $ for generic 
angles $\, \alpha \, $.
 
\vspace{1mm}

 For ``magic'' angles $\, \alpha \, $ the potentials 
$\, V (x, y, \alpha , {\bf a}) \, $ are periodic, 
and each of them has its own interval 
$\, \left[ \hat{c}_{1} (\alpha, {\bf a}) , \, 
\hat{c}_{2} (\alpha, {\bf a}) \right] \, $, 
so that
$$\left[ \hat{c}_{1} (\alpha, {\bf a}) , \, 
\hat{c}_{2} (\alpha, {\bf a}) \right] \,\,\, \subset \,\,\,
\left[ c_{1} (\alpha) , \, c_{2} (\alpha) \right] $$

 For such potentials, the presence of the situation 
$\, A_{-} \, $ in the plane $\, \Pi \, $ means 
$\, c < \hat{c}_{1} (\alpha, {\bf a}) \, $ 
(but not necessarily $\, c < c_{1} (\alpha) \, $). 
Similarly, the presence of the situation $\, A_{+} \, $ 
in the plane $\, \Pi \, $ means in this case 
$\, c > \hat{c}_{2} (\alpha, {\bf a}) \, $ 
(but not necessarily $\, c > c_{2} (\alpha) \, $).

\vspace{2mm}

 As we have already said, the potentials 
$\, V (x, y, \alpha , {\bf a}) \, $ represent a very 
special class of quasiperiodic functions on the plane. 
In particular, for each of the potentials 
$\, V (x, y, \alpha , {\bf a}) \, $, corresponding 
to a ``non-magic'' rotation angle $\, \alpha \, $, 
we have here
$$ c_{1} (\alpha) \, = \, c_{2} (\alpha) \, = \, c_{0} (\alpha)  $$

 As we also have already noted, we will be particularly 
interested here in the behavior of the constant 
$\, G (c) \, $ for $\, c \rightarrow c_{0} (\alpha) \, $ 
for such potentials.

\section{Features of periodic potentials 
$\, V (x, y, \alpha , {\bf a}) \, $ }
\setcounter{equation}{0}

  In our consideration, an important role will be 
played by periodic potentials 
$\, V (x, y, \alpha , {\bf a}) \, $, 
arising at ``magic'' rotation angles $\, \alpha \, $. 
For 6-order symmetry, magic angles $\, \alpha \, $ were 
described in detail in \cite{Shallcross1,Shallcross2}. 
Here we give a somewhat simplified description of them, 
which we also use for the case of 3-order symmetry.

 For both the 3rd order symmetry potentials and 
the 6th order symmetry potentials, the lattice of 
periods is regular triangular. Let us choose the basis 
vectors of the potential $\, V_{1} \, $ in the form
\begin{equation}
\label{e1e2}
{\bf e}_{1} \,\,\, = \,\,\, \big(T, \, 0 \big) 
\,\,\, , \quad {\bf e}_{2} \,\,\, = \,\,\, 
\big( T/2 , \, \sqrt{3} T / 2 \big) 
\end{equation}

\begin{figure}[t]
\begin{center}
\includegraphics[width=\linewidth]{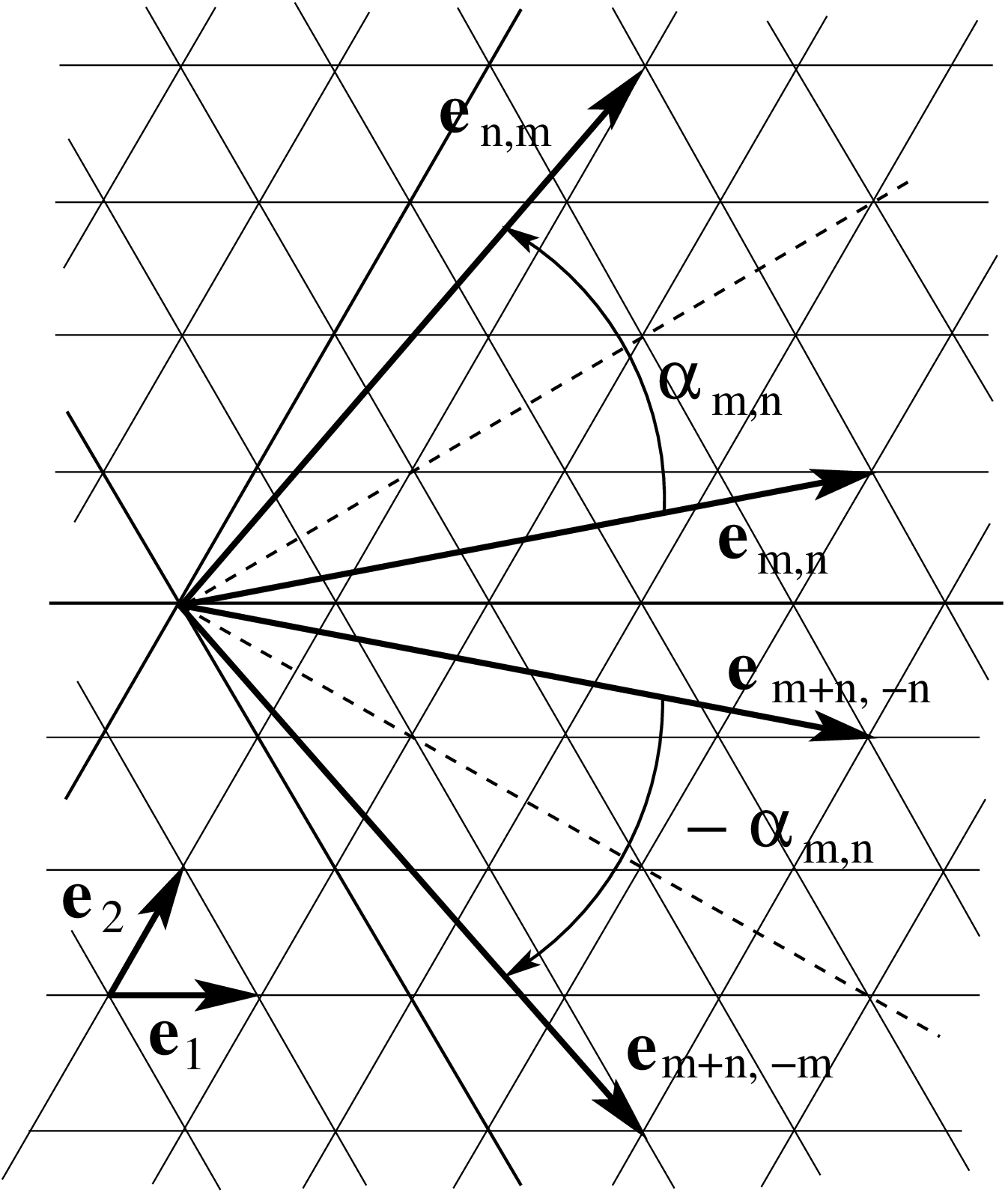}
\end{center}
\caption{``Magic'' rotation angles that translate 
the vector
$m \, {\bf e}_{1} \, + \, n \, {\bf e}_{2}$ 
into the vector
$n \, {\bf e}_{1} \, + \, m \, {\bf e}_{2}$,
and the vector
$\, (m + n) \, {\bf e}_{1}\, - \, n \, {\bf e}_{2} \, $ 
into the vector
$\, (m + n) \, {\bf e}_{1}\, - \, m \, {\bf e}_{2} \, $ 
in the triangular lattice.} 
\label{ThirdOrderPovmn}
\end{figure}

 Let us define a series of ``magic'' angles 
$\, \alpha_{m,n} \, $ using rotation from the vector
$${\bf e}_{m,n} \,\,\, = \,\,\, m \, {\bf e}_{1} \,\, + \,\, 
n \, {\bf e}_{2} $$
to the vector
$${\bf e}_{n,m} \,\,\, = \,\,\, n \, {\bf e}_{1} \,\, + \,\, 
m \, {\bf e}_{2} \,\,\, , $$
where $\, m \, $ and $\, n \, $ are relatively prime 
and $\, m > n > 0 \, $ (Fig. \ref{ThirdOrderPovmn}).

 As is easy to see, the angles $\, \alpha_{m,n} \, $ 
belong to the interval $\, (0, \, \pi / 3 ) \, $. 
In the case of the sixth order symmetry, they 
define a family of ``magic'' angles that is sufficient 
for us. In the case of the third order symmetry, we need 
to add also the angles $\, - \, \alpha_{m,n} \, $ 
for $\, m > n > 0 \, $ (Fig. \ref{ThirdOrderPovmn}).

\vspace{1mm}

\noindent
Comment.

 The potentials $\, V (x, y, - \alpha , {\bf a}) \, $, 
in fact, coincide (up to a rotation) with the potentials 
$\, \widehat{V} (x, y, \alpha , {\bf a}) \, $, 
corresponding to the replacement of the potential 
$\, V_{1} (x, y) \, $ by $\, \overline{V}_{1} (x, y) \, $. 
For many purposes, therefore, we need only consider here 
the interval $\, \alpha \in (0, \, \pi / 3 ) \, $ also in 
the general case (third-order symmetry).

\vspace{1mm}

 It is easy to see that all potentials 
$\, V (x, y, \alpha_{m,n}, {\bf a}) \, $, as well as 
$\, V (x, y, - \alpha_{m,n}, {\bf a}) \, $, are periodic. 
The minimal periods for the potentials 
$\, V (x, y, \alpha_{m,n}, {\bf a}) \, $ 
are either the vectors
$${\bf b}_{1} \,\,\, = \,\,\, {\bf e}_{m+n, -n} \,\,\, , 
\quad \quad {\bf b}_{2} \,\,\, = \,\,\, {\bf e}_{n,m}  $$
(if $\, m - n \neq 3k \, $, $\,\, k \in \mathbb{N} $),
or
$${\bf b}^{\prime}_{1} \,\,\, = \,\,\, 
(n + 2 k) \, {\bf e}_{1} \,\, - \,\, (n + k) \, {\bf e}_{2} 
\,\,\, , $$
$${\bf b}^{\prime}_{2} \,\,\, = \,\,\, 
(n + k)  \, {\bf e}_{1}  \,\, + \,\, k \, {\bf e}_{2} $$
(if $\, m - n = 3k \, $, $\,\, k \in \mathbb{N} $). 

\vspace{2mm}

 The length of the minimal periods of 
$\, V (x, y, \alpha_{m,n}, {\bf a}) \, $ is equal to 
$\, \sqrt{m_{0}^{2} \, + \, n_{0}^{2} \, + \, m_{0} \, n_{0}} \,\, $, 
where
$$(m_{0}, n_{0}) \,\,\, = \,\,\, (m, n) \,\,\, , \quad
m - n \, \neq \, 3 k \,\,\, , \,\,\,\,\, k \in \mathbb{N} $$
\begin{multline*}
(m_{0}, n_{0}) \,\,\, = \,\,\,  (n + k, \, k ) \,\,\, , \quad
m - n \, = \, 3 k \,\,\, , \,\,\,\,\, k \in \mathbb{N} 
\end{multline*}

 Similarly, the minimal periods for the potentials 
$\, V (x, y, - \alpha_{m,n}, {\bf a}) \, $ are either vectors
$${\bf b}_{1} \,\,\, = \,\,\, {\bf e}_{m+n, -m} \,\,\, , \quad
{\bf b}_{2} \,\,\, = \,\,\, {\bf e}_{m, n} \,\,\, , $$
(if $\, m - n \neq 3k \, $, $\,\, k \in \mathbb{N} $),
or
$${\bf b}^{\prime}_{1} \,\,\, = \,\,\, 
(n + 2 k) \, {\bf e}_{1} \,\, - \,\, k \, {\bf e}_{2} 
\,\,\, , $$ 
$${\bf b}^{\prime}_{2} \,\,\, = \,\,\, 
k  \, {\bf e}_{1}  \,\, + \,\, (n + k) \, {\bf e}_{2} $$
(if $\, m - n = 3k \, $, $\,\, k \in \mathbb{N} $),
having the same length.

\vspace{1mm}

 The value of the angle $\, \alpha_{m,n} \, $ 
is determined by the relation
$$\tan \alpha_{m,n} \,\, = \,\, 
{\sqrt{3} \, (m^{2} - n^{2}) \over m^{2} + n^{2} + 4 m n} 
\,\, = \,\, {\sqrt{3} \, ( 1 - (n/m)^{2}) \over 
(n/m)^{2} + 4 (n/m) + 1} $$

 It is easy to verify that in the region 
$\, 0 < \alpha < 60^{\circ} \, $ the relation
$$\tan \alpha \,\,\, = \,\,\, 
{\sqrt{3} \, ( 1 - w^{2} ) \over w^{2} + 4 w + 1} $$
implies
$$0 \,\,\, < \,\,\, w \,\,\, = \,\,\, {\sqrt{3} 
\,\, - \,\, \tan \, \alpha \over
\sqrt{3} \, \sqrt{\tan^{2} \alpha + 1} 
\,\, + \,\, 2 \, \tan \, \alpha } \,\,\, < \,\,\, 1 $$

 The derivative of the function
$$f (w) \,\,\ = \,\,\,  {\sqrt{3} \, ( 1 - w^{2} ) \over 
w^{2} + 4 w + 1} $$
has the form
$${d f \over d w}  \,\,\ = \,\,\, - \, 4 \sqrt{3} \,\, 
{w^{2} + w  + 1 \over (w^{2} + 4 w + 1)^{2}} $$
and, thus, in the region $\, 0 < w < 1 \, $
$$\left| { d f \over d w} \right| \,\,\, < \,\,\, 
{4 \sqrt{3} \over w^{2} + 4 w + 1} \,\,\, < \,\,\, 
4 \, \sqrt{3} $$

 Using Lagrange's theorem, we can then write for $m > n$
 $$\left| {n \over m} - w \right| \,\, < \,\, {1 \over m^{2}} 
\,\,\, \Rightarrow \,\,\, 
\left| \tan \, \alpha_{m,n} \, - \, \tan \, \alpha \right| 
\,\, < \,\, {4 \, \sqrt{3}  \over m^{2}} \,\,\, \Rightarrow  $$
$$\Rightarrow \,\,\, \left| \alpha_{m,n} \, - \, \alpha \right| 
\,\,\, < \,\,\, 4 \, \sqrt{3} \,\, {1 \over m^{2}} $$

 As is well known, any number $\, w \, $ has an infinite 
sequence of ``good'' approximations by rational numbers 
$\, n^{(s)} / m^{(s)} \, $ with increasing $\, m^{(s)} \, $, 
such that
\begin{equation}
\label{wPribl}
\left| {n^{(s)} \over m^{(s)}} - w \right| \,\, < \,\, 
{1 \over \left(m^{(s)}\right)^{2}} 
\end{equation}

 We can then conclude that each of the generic angles 
$\, 0 < \alpha < 60^{\circ} \, $ also has an infinite 
sequence of ``good'' approximations by ``magic'' angles
$\, \alpha_{m^{(s)},n^{(s)}} \, $,
such that
\begin{equation}
\label{GenAnglePribl}
\left| \alpha_{m^{(s)},n^{(s)}} \, - \, \alpha \right| 
\,\,\, < \,\,\, 4 \, \sqrt{3} \,\, 
{1 \over \left(m^{(s)}\right)^{2}}
\end{equation}

\vspace{1mm}

\begin{figure}[t]
\begin{center}
\includegraphics[width=\linewidth]{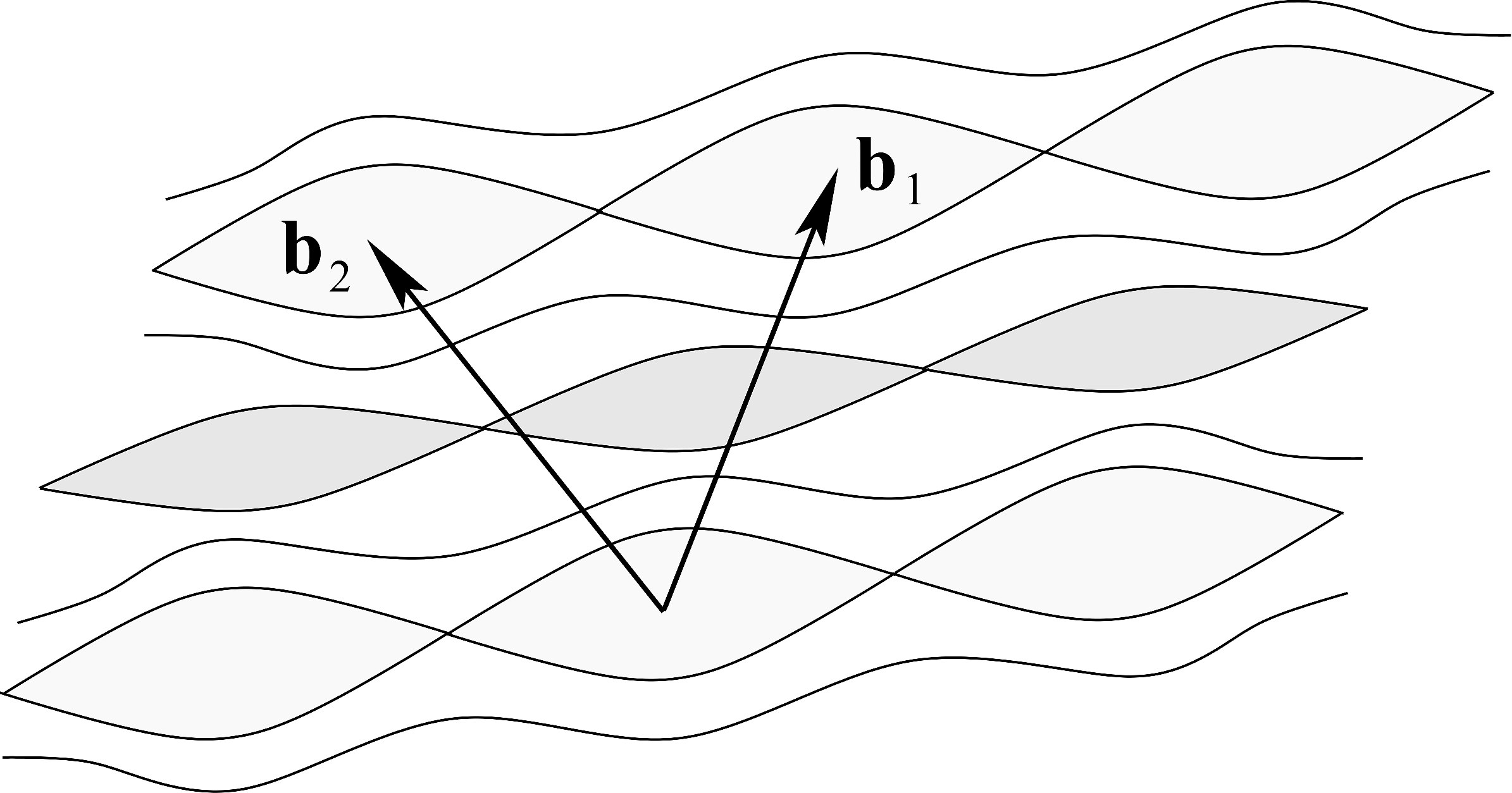}
\end{center}
\caption{Periodic level lines of potential 
$\, V (x, y, \, \alpha_{m,n} , \, {\bf a}) \, $ 
(schematically)}
\label{PerLevLines}
\end{figure}

 Open level lines of potentials 
$\, V (x, y, \alpha_{m,n}, {\bf a}) \, $, corresponding 
to finite intervals 
$\, \left[ \hat{c}_{1} (\alpha_{m,n}, {\bf a}), \, 
\hat{c}_{2} (\alpha_{m,n}, {\bf a}) \right] \, $ 
($\hat{c}_{2} > \hat{c}_{1}$), are periodic and have 
some common integer direction in the basis 
$\, \{ {\bf b}_{1}, \, {\bf b}_{2} \} \, $ or 
$\, \{ {\bf b}^{\prime}_{1}, \, {\bf b}^{\prime}_{2} \} \, $ 
(Fig. \ref{PerLevLines}). These directions, however, 
can be different for different values of
the parameters $\, (a^{1}, a^{2}) \, $.

 As shown in \cite{Superpos}, the lengths of the intervals
$\, \left[ \hat{c}_{1} (\alpha_{m,n}, {\bf a}), \,
\hat{c}_{2} (\alpha_{m,n}, {\bf a}) \right] \, $ and
$\, \left[ c_{1} (\alpha_{m,n}), \,
c_{2} (\alpha_{m,n}) \right] \, $ cannot exceed the values
$\, C_{1} T / \sqrt{3 (m_{0}^{2} + n_{0}^{2} + m_{0} n_{0})} \, $
and
$\, 2 C_{1} T / \sqrt{3 (m_{0}^{2} + n_{0}^{2} + m_{0} n_{0})} \, $
respectively. 

\vspace{1mm}

\begin{figure}[t]
\begin{center}
\includegraphics[width=\linewidth]{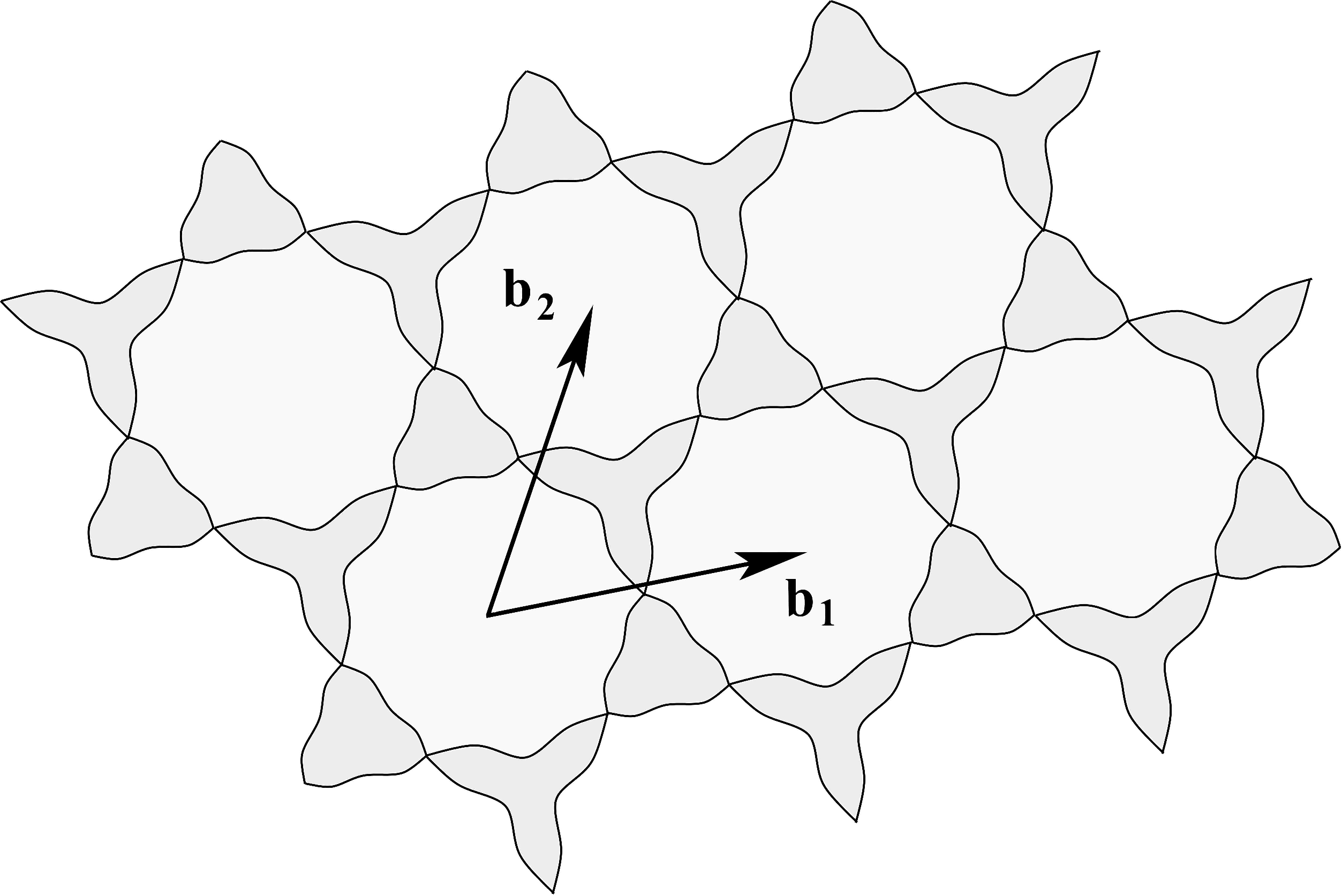}
\end{center}
\caption{Example of a ``singular net'' for a potential 
$\, V (x, y, \, \alpha_{m,n} , \, 0, 0) \, $ 
with non-degenerate critical points}
\label{SingNet}
\end{figure}

 Potentials $\, V (x, y, \, \alpha_{m,n} , \, 0, 0) \, $ 
have rotational symmetry with respect to the origin and 
obviously cannot have non-singular periodic open level 
lines. The situations $\, A_{-} \, $ and $\, A_{+} \, $ 
are separated here by a ``singular'' periodic net 
arising at a single level 
$\, c = c_{0} (\alpha_{m,n}) \, $ (Fig. \ref{SingNet}). 
The centers of rotational symmetry of the potential 
$\, V (x, y, \, \alpha_{m,n} , \, 0, 0) \, $, 
as is not difficult to see, are located at the points
\begin{equation}
\label{SymCentpq}
p \,\, {2 {\bf b}_{1} - {\bf b}_{2} \over 3} \,\,\, + \,\,\,
q \,\, {{\bf b}_{1} + {\bf b}_{2} \over 3} \,\,\, , \quad \quad
p, q \, \in \, \mathbb{Z} 
\end{equation}
(or the same with the replacement of 
$\, \{{\bf b}_{1}, \, {\bf b}_{2} \} \, $ with 
$\, \{{\bf b}^{\prime}_{1}, \, {\bf b}^{\prime}_{2} \} $).

 The ``singular net'' shown in Fig. \ref{SingNet} 
corresponds to the ``generic'' case, in particular, 
all saddle points of the potential $\, V (x, y) \, $ 
are non-degenerate here, and the ``net'' contains 
exactly 3 nonequivalent saddle points. It is easy 
to see that all ``cells'' of such a net have a rotational 
symmetry. This property holds, in fact, for any generic 
``singular net'' in our situation. In a more general 
case, the ``singular net'' can be more complicated, 
however, this requires a special construction of 
the corresponding potential $\, V (x, y) \, $ 
(in particular, we can assume that such situations 
do not arise in ``real'' two-layer systems).
 
\vspace{1mm}

 As we have already said, the potentials 
$\, V (x, y, \, \alpha, \, {\bf a}) \, $ are given 
by the superposition of the potential 
$\, V_{1} (x, y) \, $ and the potential 
$\, V_{2} (x, y) \, $, which has the form (\ref{V2Pot}). 
In addition, we (without loss of generality) assume 
that the basis vectors of the lattice of periods of 
the potential $\, V_{1} \, $ are given by 
the relations (\ref{e1e2}). It is easy to see that 
the potential $\, \overline{V}_{1} (x, y) \, $ has 
the same lattice of periods, and the periods of 
the potential $\, V_{2} (x, y) \, $ 
are given by the vectors
$$\pi_{\alpha} ({\bf e}_{1}) \,\,\, ,  \quad 
\pi_{\alpha} ({\bf e}_{2}) $$

 It is easy to see then that
$$V \big( 
{\bf r}, \, \alpha , \,\, {\bf a} + \pi_{\alpha} ({\bf e}_{1,2}) \big)
\,\,\, \equiv \,\,\, V ({\bf r}, \, \alpha, \, {\bf a} ) $$
and
$$V \big( 
{\bf r}, \, \alpha , \,\, {\bf a} + {\bf e}_{1,2} \big)
\,\,\, \equiv \,\,\, V \big( {\bf r} - {\bf e}_{1,2}, \,\, 
\alpha, \, {\bf a} \big) $$

 The potentials 
$\, V \big( {\bf r}, \, \alpha , \, 
{\bf a} + {\bf e}_{1,2} \big) $ and 
$\, V ({\bf r}, \, \alpha, \, {\bf a} ) \, $
can be considered equivalent, since they 
differ only in a shift in the plane $\, (x, y) $. 
In general, for each potential 
$\, V ({\bf r}, \, \alpha , \, {\bf a}) \, $ one can define 
a whole class of potentials equivalent to it, setting
\begin{multline*}
V_{klpq} \big( {\bf r}, \, \alpha, \, {\bf a} \big) 
\,\,\, =   \\
= \,\,\, V \big( {\bf r}, \, \alpha, \,\, 
{\bf a} + k \pi_{\alpha} ({\bf e}_{1}) + 
l \pi_{\alpha} ({\bf e}_{2}) + p {\bf e}_{1} + q {\bf e}_{2} \big) 
\end{multline*}
($k, l, p, q \, \in \, \mathbb{Z}$).

\vspace{2mm}

 For generic (not ``magic'') angles $\, \alpha \, $ 
any set of vectors
$${\bf a}_{0} \, + \, k \, \pi_{\alpha} ({\bf e}_{1})  +  
l \, \pi_{\alpha} ({\bf e}_{2}) + p \, {\bf e}_{1} + 
q \, {\bf e}_{2} \,\, ,  \quad
k, l, p, q \, \in \, \mathbb{Z} $$
is everywhere dense in the space of the parameters 
$\, {\bf a} \, $. 

\vspace{2mm}
 
 For angles $\, \alpha_{m,n} \, $ vectors
$$k \, \pi_{\alpha} ({\bf e}_{1}) \, + \, 
l \, \pi_{\alpha} ({\bf e}_{2}) \, + \, p \, {\bf e}_{1} 
\, + \, q \, {\bf e}_{2} $$
form a triangular lattice with step 
$\, T / \sqrt{m_{0}^{2} + n_{0}^{2} + m_{0} n_{0}} \, $.
In particular, each potential 
$\, V (x, y, \, \alpha_{m,n}, \, {\bf a}) \, $ 
is equivalent to some potential 
$\, V (x, y, \, \alpha_{m,n}, \, {\bf a}^{\prime}) \, $ 
such that 
$\, |{\bf a}^{\prime}| \, \leq \, 
T / \sqrt{3 (m_{0}^{2} + n_{0}^{2} + m_{0} n_{0})} \, $.

\vspace{1mm}

 Obviously, equivalent potentials have level lines of 
the same type at each level $\, c \, $. In particular, 
all potentials equivalent to 
$\, V (x, y, \, \alpha_{m,n} , \, 0, 0) \, $ 
have a ``singular net'' at the same level 
$\, c_{0} (\alpha_{m,n}) \, $.

\vspace{1mm}

 In our case, all potentials 
$\, V (x, y, \, \alpha_{m,n}, \, {\bf a}) \, $, 
equivalent to the potential 
$\, V (x, y, \, \alpha_{m,n} , \, 0, 0) \, $, 
also have reflection symmetry. Their axes of symmetry, 
as is easy to see, form the angles
$${1 \over 2} \, \alpha_{m,n} \,\,\, , \quad \quad
{1 \over 2} \, \alpha_{m,n} \, + \, {2\pi \over 3} 
\quad \quad \text{and} \quad \quad
{1 \over 2} \, \alpha_{m,n} \, + \, {4\pi \over 3} $$
with the vector $\, {\bf e}_{1} \, $.

 In particular, all such lines passing through the points 
(\ref{SymCentpq}) represent the symmetry axes of the 
potential $\, V (x, y, \, \alpha_{m,n} , \, 0, 0) \, $.

\vspace{1mm}

 We will also need the following lemma here.
 
\vspace{2mm}

\noindent
{\bf Lemma 3.1}

 Let a doubly periodic potential $\, V (x, y) \, $ 
have a third-order rotation symmetry and a reflection 
symmetry. Then there exists a universal constant 
$\, D \, $ such that the diameter of any bounded 
connected component of the set
\begin{equation}
\label{Vlessc}
V (x, y) \,\,\, < \,\,\, c \quad \quad (\forall c) \,\,\, ,
\end{equation}
as well as
\begin{equation}
\label{Vgreatc}
V (x, y) \,\,\, > \,\,\, c
\end{equation}
does not exceed $\, D L \, $, where $\, L \, $ is the 
length of the periods of the potential $\, V (x, y) \, $.

\vspace{2mm}

\noindent
Proof.

 To prove the lemma, it is obviously sufficient to prove 
its assertion for the cells of the ``singular net'' 
$\, V (x, y) = c_{0} \, $, since no boundaries of the 
sets (\ref{Vlessc}) and (\ref{Vgreatc}) can intersect it.

 We will give a proof for the generic case, when the 
potential $\, V (x, y) \, $ has a ``standard'' singular net 
at the level $\, c_{0} \, $ (e.g. Fig. \ref{SingNet}), and 
each cell of the singular net has rotational symmetry 
as well as reflection symmetry. In the general case, 
we can assume that any cell of the singular net 
is included in some symmetric ``conglomerate'' that 
does not intersect its shifts, and perform an estimate 
for the entire conglomerate. The corresponding 
``conglomerate'' can be determined (ambiguously), 
for example, using a small symmetric perturbation 
of the potential $\, V (x, y) \, $ that leads to the 
generic situation.

\begin{figure}[t]
\begin{center}
\includegraphics[width=0.9\linewidth]{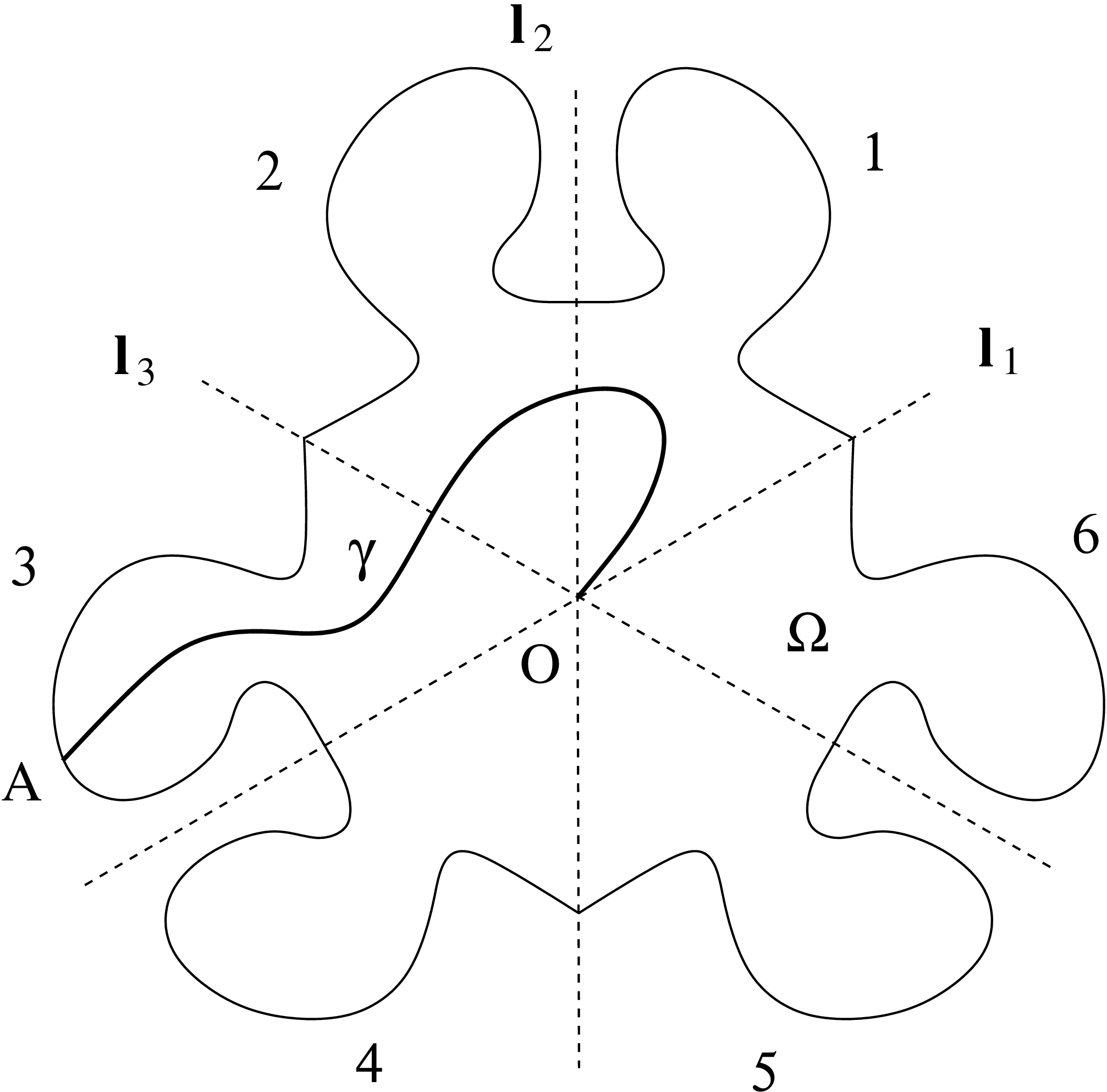}
\end{center}
\caption{A symmetric cell of a ``singular net'' 
$\, V (x, y) = c_{0} \, $ and a path $\, \gamma \, $ 
connecting its center with the most distant 
point on the boundary.}
\label{SymOmega}
\end{figure}

 Let, therefore, a cell of a singular net be a symmetric 
(simply connected) region $\, \Omega \, $ with a boundary 
$\, \partial \Omega \, $ formed by singular level lines 
$\, V (x, y) = c_{0} \, $ (Fig. \ref{SymOmega}).

 Let $\, \gamma \subset \Omega \, $ be some curve 
connecting the center $\, O \, $ of the domain 
$\, \Omega \, $ with the most distant point $\, A \, $ 
on its boundary. We can assume that $\, \gamma \, $ 
passes through the point $\, O \, $ only once, and also 
intersects the lines $\, l_{1} \, $, $\, l_{2} \, $ and 
$\, l_{3} \, $ only finitely many times 
(the latter can be seen by considering $\, \gamma \, $ 
as generic finite-link broken line). Obviously, 
the distance between the endpoints of $\, \gamma \, $ 
is not less than $\, d/2 \, $, where $\, d \, $ is 
the diameter of the domain $\, \Omega \, $.

 Through the center of the region $\, \Omega \, $ there 
pass 3 axes of symmetry ($\, l_{1} \, $, $\, l_{2} \, $ 
and $\, l_{3} \, $), dividing the plane into 6 sectors. 
Let the initial direction of the curve $\, \gamma \, $ 
lie in sector 1 (Fig. \ref{SymOmega}).

\begin{figure}[t]
\begin{center}
\includegraphics[width=\linewidth]{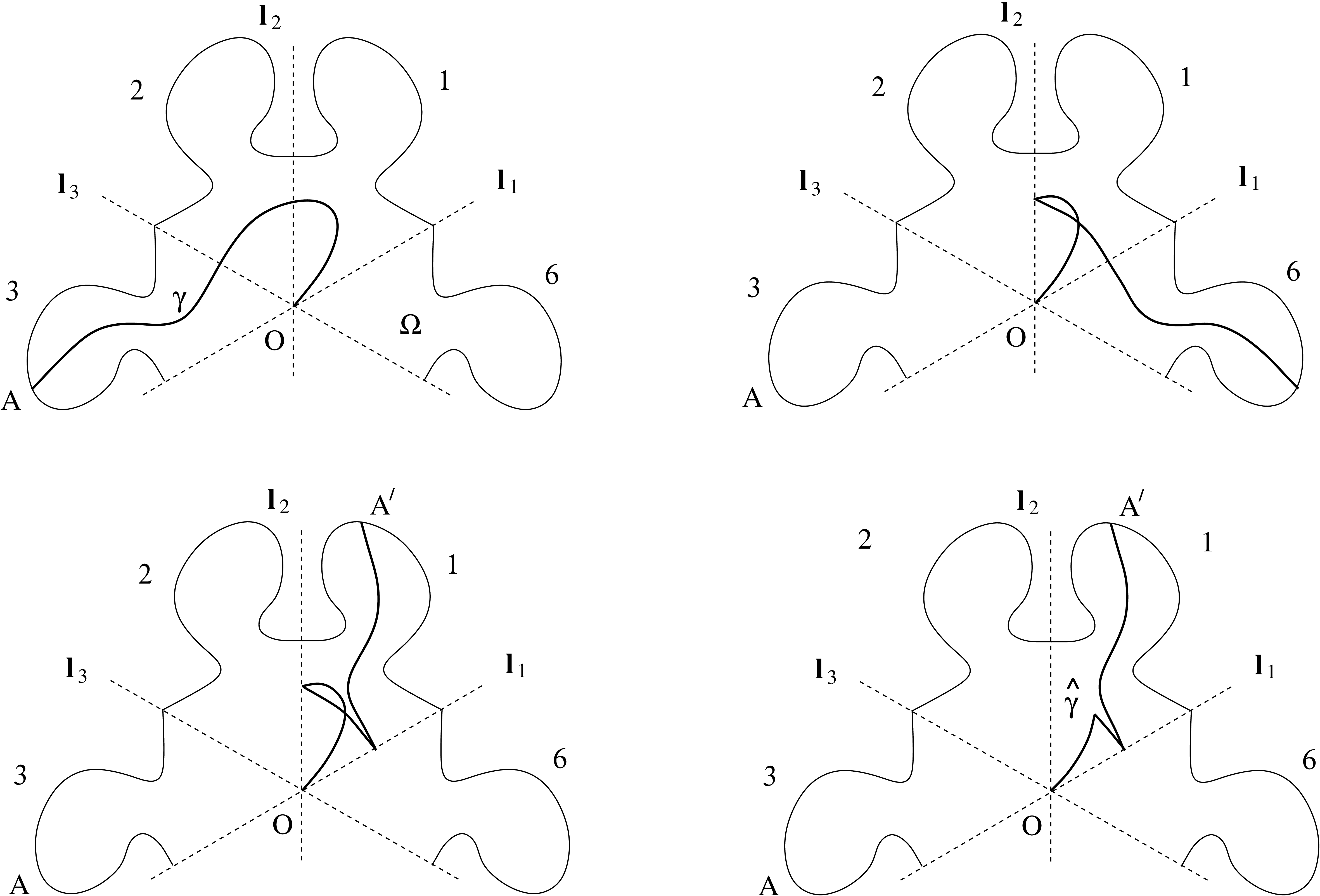}
\end{center}
\caption{The curve $\, \widehat{\gamma} \, $ obtained from 
$\, \gamma \, $ by a finite number of reconstructions and 
lying entirely in sector 1 inside $\, \Omega \, $.}
\label{hatgamma}
\end{figure}

 Using reflections with respect to the lines $\, l_{1} \, $ 
and $\, l_{2} \, $, as well as reconstructions of the curve 
$\, \gamma \, $, we can construct a non-self-intersecting 
curve $\, \widehat{\gamma} \subset \Omega \, $ that lies 
entirely in sector 1 and connects the point $\, O \, $ with 
the point $\, A^{\prime} \, $, which is at the same 
distance from the center of $\, \Omega \, $ as the point 
$\, A \, $ (Fig. \ref{hatgamma}).

 It is also easy to see that with a small perturbation 
we can achieve that $\, \widehat{\gamma} \, $ is a smooth 
curve with interior lying inside sector 1 
($0 < \varphi < 60^{\circ}$).
 
\begin{figure}[t]
\begin{center}
\includegraphics[width=0.8\linewidth]{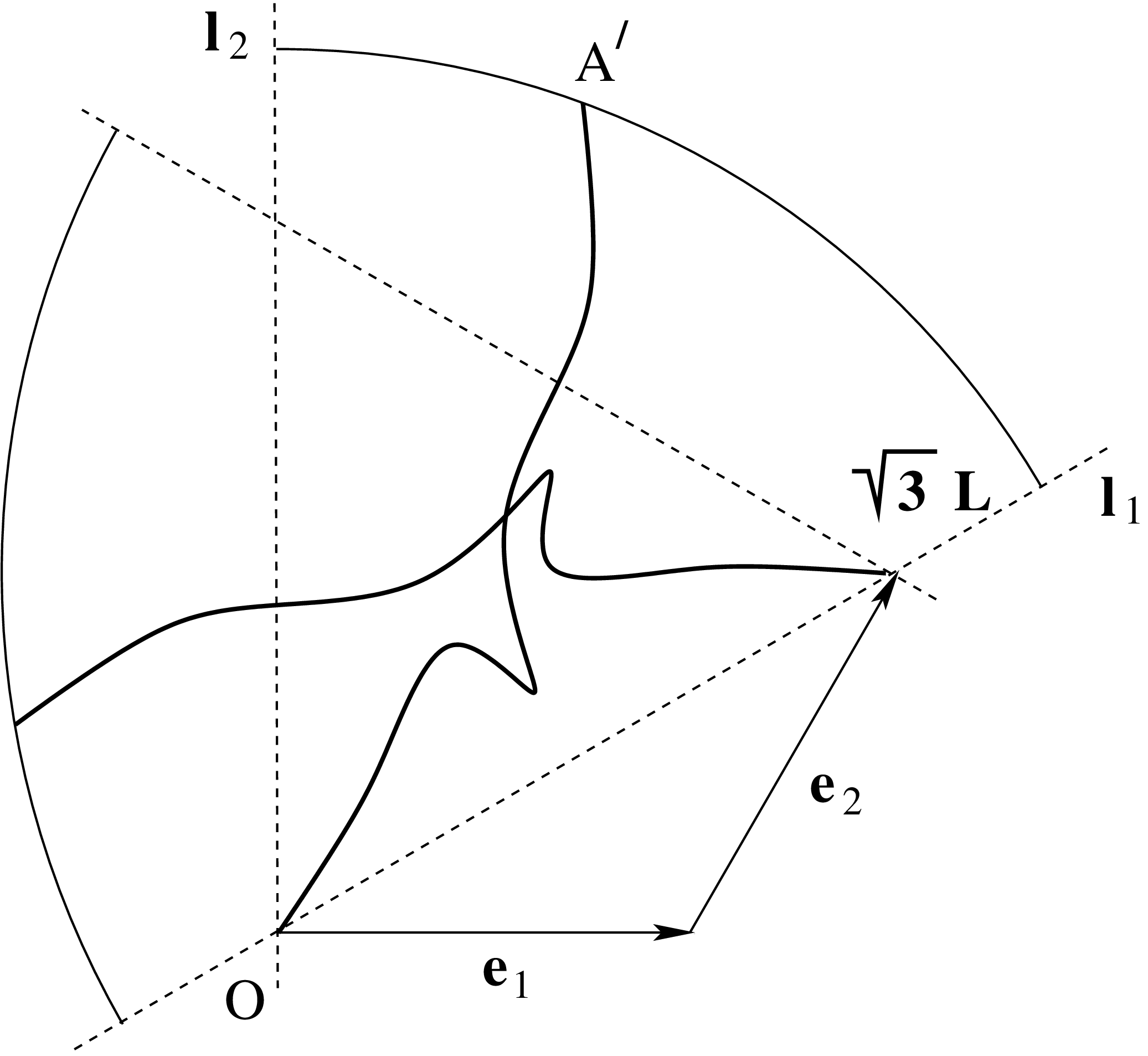}
\end{center}
\caption{The intersection of $\, \widehat{\gamma} \, $ 
and the shift of 
$\, \pi_{120^{\circ}} \left( \widehat{\gamma} \right) \, $ 
by a period of $\, V (x, y) \, $ for 
$\, | O A^{\prime} | \, > \, \sqrt{3} L \, $. }
\label{gammashiftgamma}
\end{figure}

 The curve $\, \widehat{\gamma} \, $ obviously does not 
intersect at interior points with its rotations by 
$\, 120^{\circ} \, $ relative to the centers of symmetry 
and shifts by periods of the potential $\, V (x, y) \, $. 
Considering its rotation by $\, 120^{\circ} \, $ relative 
to the point $\, O \, $ and shift by period $\, V (x, y) \, $ 
along the line $\, l_{1} \, $, one can see, however, that 
this is possible only if the distance between the ends of 
$\, \widehat{\gamma} \, $ does not exceed $\, \sqrt{3} L \, $ 
(Fig. \ref{gammashiftgamma}).

 In particular, it follows that for the constant 
$\, D \, $ one can use the value 
$\, D \, = \, 2 \sqrt{3} \, $ 
(this estimate can be significantly improved, here, 
however, we do not aim to obtain an exact estimate 
for the constant $\, D $).

\hfill{Lemma 3.1 is proven.}

\section{Level lines of quasi-periodic potentials 
$\, V (x, y, \alpha, {\bf a}) \, $ }
\setcounter{equation}{0}

 In this section we consider potentials 
$\, V (x, y, \alpha, {\bf a}) \, $ arising 
for ``generic'' (not ``magic'') angles $\, \alpha \, $. 
As we have already noted, such potentials are quasiperiodic 
functions in $\, \mathbb{R}^{2} \, $ with 4 quasiperiodes.

 For a given value of $\, \alpha \, $, open level lines
of potentials 
$$V (x, y, \alpha, {\bf a}) \,\,\, , \quad 
{\bf a} \in \mathbb{R}^{2} $$
arise here (at least for one value of 
$\, {\bf a}$) only at a single level 
$\, c = c_{0} (\alpha) \, $ (\cite{Superpos}).
Each specific potential $\, V (x, y, \alpha, {\bf a}) \, $ 
has either open level lines or closed level lines of 
arbitrarily large sizes (or both), at the level 
$\, c_{0} (\alpha) \, $ (see \cite{DynMalNovUMN}). 
In any case, the maximum size of connected level lines 
of each $\, V (x, y, \alpha, {\bf a}) \, $ at the level 
$\, c_{0} (\alpha) \, $ turns into infinity.

 As we have already noted, for 
$\, c \neq c_{0} (\alpha) \, $ the sizes of all connected 
level lines of potentials $\, V (x, y, \alpha, {\bf a}) \, $ 
are limited by one constant $\, G (c) \, $. Here we want 
to obtain the estimate
\begin{equation}
\label{CEstim}
G (c) \,\,\, \leq \,\,\, C \,\, \left| c - c_{0} (\alpha) 
\right|^{-1} 
\end{equation}
(for some constant $\, C$), for ``almost all'' potentials 
$\, V (x, y, \alpha, {\bf a}) \, $, using an approximation 
of quasiperiodic potentials $\, V (x, y, \alpha, {\bf a}) \, $ 
by the potentials $\, V (x, y, \alpha_{m,n} , {\bf a}) \, $ 
described in the previous section. 

 It is easy to see that the estimate (\ref{CEstim}) 
is a special case of the more general relation
$$G \,\,\, \sim \,\,\,  \left| c - c_{0} \right|^{-\nu} 
\,\,\, , $$
that arises when describing, for example, level lines 
of random potentials. It may be noted here that the 
value $\, \nu = 1 \, $ differs, for example, from the
value $\, \nu = 4/3 \, $ and other values arising in
the percolation theory in two-dimensional systems 
(see, for example, \cite{Stauffer, Essam, Riedel, Trugman}). 
This circumstance is one of the differences between 
the potentials considered here and the more generally 
accepted models of random potentials on the plane. 
In general, it may be noted, however, that the use of 
quasiperiodic potentials as a model of random potentials, 
both in the case of one-dimensional and in the case of 
two-dimensional systems, is also widespread 
(see, for example, \cite{WZLWLMHXMCJ, LesserLifshitz} 
and the references therein).

\vspace{1mm}

 Let us prove the following lemma here.
 
\vspace{1mm}

\noindent
{\bf Lemma 4.1}

 For every ``non-magic'' angle $\, \alpha \, $ 
there is a sequence $\, \Delta_{(s)} > 0 \, $, 
$\, \Delta_{(s)} \rightarrow 0 \, $, 
$\, s \rightarrow \infty \, $, such that
$$\forall c \,\,\, \notin \,\,\, \left[ 
c_{0} (\alpha) - \Delta_{(s)} \, , \,\, 
c_{0} (\alpha) + \Delta_{(s)} \right] $$
the size of connected components of level lines
$$V (x, y, \alpha, {\bf a}) \,\,\, = \,\,\, c $$
does not exceed
$$C \, \Delta_{(s)}^{-1} $$
(with some constant $\, C$ that is universal for 
the family $\, V (x, y, \alpha, {\bf a}) \, $).

\vspace{1mm}

\noindent
Proof.

 Since each closed level line 
$\, V (x, y, \alpha, {\bf a}) \, = \, c \, $ 
belongs to the boundary of a bounded region
\begin{equation}
\label{VminusComp}
V (x, y, \alpha, {\bf a}) \,\,\, < \,\,\, c 
\end{equation}
or
\begin{equation}
\label{VplusComp}
V (x, y, \alpha, {\bf a}) \,\,\, > \,\,\, c \,\,\, ,
\end{equation}
it suffices to prove the assertion of the lemma 
for at least one of the regions (\ref{VminusComp}) 
or (\ref{VplusComp}) adjacent to it.

 According to the comment made in the previous section,
we will also assume here $\, 0 < \alpha < 60^{\circ} \, $.

 Let, for example, $\, c < c_{0} (\alpha) \, $. 
Consider an arbitrary connected component of the 
set (\ref{VminusComp}) containing some point 
$\, (x_{0}, y_{0}) \, $.

 Consider a sequence of ``magic'' numbers
$$\alpha^{(s)} \,\,\, = \,\,\, \alpha_{m^{(s)},n^{(s)}} 
\,\,\, , $$
that give ``good'' approximations to the angle 
$\, \alpha \, $, such that, 
according to (\ref{GenAnglePribl})
$$\left| \alpha^{(s)} \, - \, \alpha \right| 
\,\,\, < \,\,\, 4 \, \sqrt{3} \,\, 
{1 \over \left(m^{(s)}\right)^{2}} \,\,\, , $$
$$m^{(s)} \rightarrow \infty \,\,\, , \quad 
n^{(s)} \rightarrow \infty \,\,\, , \quad \quad
s \rightarrow \infty $$

 Consider the potential
$\, \widetilde{V}^{(s)}_{(x_{0},y_{0})} 
(x, y, \alpha, {\bf a}) \, $,  
defined by the superposition of $\, V_{1} (x, y) \, $ 
and the potential
$\, \widetilde{V}^{(s)}_{2 (x_{0},y_{0})} 
(x, y, \alpha, {\bf a}) \, $, obtained by rotating of 
$\, V_{2} (x, y) \, $ around the point 
$\, (x_{0}, y_{0}) \, $ by the angle
$$\delta \alpha^{(s)} \,\,\, = \,\,\, \alpha^{(s)} 
\,\, - \,\, \alpha $$

 Obviously,
$$\widetilde{V}^{(s)}_{(x_{0},y_{0})} 
(x, y, \alpha, {\bf a}) \,\,\, = \,\,\, 
V \left( x, y, \, \alpha^{(s)} , \, {\bf a}^{\prime} 
\right) $$
for some $\, {\bf a}^{\prime} \, $ and is a periodic 
potential with periods
$${\bf b}_{1} \,\,\, = \,\,\, 
\left( m^{(s)} + n^{(s)} \right) \, {\bf e}_{1} \,\, - \,\,
n^{(s)} \, {\bf e}_{2} \,\,\, , $$
$${\bf b}_{2} \,\,\, = \,\,\, n^{(s)} \, {\bf e}_{1} 
\,\, + \,\, m^{(s)} \, {\bf e}_{2} $$
(if $\, m^{(s)} - n^{(s)} \neq 3 k^{(s)} \, $, 
$\,\, k^{(s)} \in \mathbb{N} $),
or
$${\bf b}^{\prime}_{1} \,\,\, = \,\,\, 
\left( n^{(s)} + 2 k^{(s)} \right) \, {\bf e}_{1} 
\,\, - \,\, \left( n^{(s)} + k^{(s)} \right) \, {\bf e}_{2} 
\,\,\, , $$
$${\bf b}^{\prime}_{2} \,\,\, = \,\,\, 
\left( n^{(s)} + k^{(s)} \right)  \, {\bf e}_{1}  
\,\, + \,\, k^{(s)} \, {\bf e}_{2} $$
(if $\, m^{(s)} - n^{(s)} = 3 k^{(s)} \, $, 
$\,\, k^{(s)} \in \mathbb{N} $).

 Considering the deformation of the corresponding 
embedding (\ref{Embedding}) and using the relation 
(\ref{C1Otsen}), inside the circle of radius $\, R \, $ 
with center $\, (x_{0}, y_{0}) \, $ we obtain the relation
\begin{multline*}
\left| V (x, y, \alpha, {\bf a}) \, - \, 
\widetilde{V}^{(s)}_{(x_{0},y_{0})} (x, y, \alpha, {\bf a})
\right| \quad <  \\
< \quad C_{1} \, \delta \alpha^{(s)} \, R \quad < \quad
4 \, \sqrt{3} \,\, 
{C_{1} \, R \over \left( m^{(s)} \right)^{2} }
\end{multline*}

 As follows from the arguments of the previous section, 
there is also a potential
$$V^{(s)}_{(x_{0},y_{0})} 
(x, y, \alpha, {\bf a}) \,\,\, = \,\,\, 
V \left( x, y, \, \alpha^{(s)} , \, {\bf a}^{\prime\prime} 
\right) \,\,\, , $$
equivalent to the potential $\, V (x, y, \alpha, 0, 0) \, $ 
and such that
$$\left| {\bf a}^{\prime} \, - \, {\bf a}^{\prime\prime} 
\right| \quad \leq \quad 
{T \over \sqrt{3 \left( \left( m^{(s)}_{0} \right)^{2} + 
\left( n^{(s)}_{0} \right)^{2} + 
m^{(s)}_{0} n^{(s)}_{0} \right) } } $$

 From (\ref{C1Otsen}) we then have
\begin{multline*}
\left| 
\widetilde{V}^{(s)}_{(x_{0},y_{0})} (x, y, \alpha, {\bf a})
\, - \, V^{(s)}_{(x_{0},y_{0})} (x, y, \alpha, {\bf a})
\right| \quad \leq   \\
\leq \quad 
{C_{1} \, T \over 
\sqrt{3 \left( \left( m^{(s)}_{0} \right)^{2} + 
\left( n^{(s)}_{0} \right)^{2} + 
m^{(s)}_{0} n^{(s)}_{0} \right) } }
\end{multline*}
and, thus, in the circle of radius $\, R \, $ 
with center $\, (x_{0}, y_{0}) \, $
\begin{multline*}
\left| V (x, y, \alpha, {\bf a}) \, - \, 
V^{(s)}_{(x_{0},y_{0})} (x, y, \alpha, {\bf a})
\right| \quad <  \\
< \,\,\,
{4 \sqrt{3} \, C_{1} \, R \over 
\left( m^{(s)} \right)^{2} }
\,\,\, + \,\,\,  {C_{1} \, T \over 
\sqrt{3 \left( \left( m^{(s)}_{0} \right)^{2} + 
\left( n^{(s)}_{0} \right)^{2} + 
m^{(s)}_{0} n^{(s)}_{0} \right) } }
\end{multline*}

 The potential 
$\, V^{(s)}_{(x_{0},y_{0})} (x, y, \alpha, {\bf a}) \, $ 
has exact rotational symmetry and, according to Lemma 3.1, 
any connected component of the set
$$V^{(s)}_{(x_{0},y_{0})} (x, y, \alpha, {\bf a})
\,\,\, < \,\,\, c_{0} \left(\alpha^{(s)}, 0, 0 \right) $$
or
$$V^{(s)}_{(x_{0},y_{0})} (x, y, \alpha, {\bf a})
\,\,\, > \,\,\, c_{0} \left(\alpha^{(s)}, 0, 0 \right) $$
has the size no greater than
$$D \, L^{(s)} \,\,\, = \,\,\, D \, T \, 
\sqrt{\left( m^{(s)}_{0} \right)^{2} + 
\left( n^{(s)}_{0} \right)^{2} + 
m^{(s)}_{0} n^{(s)}_{0} } $$

 In particular, any such component containing the point 
$\, (x_{0}, y_{0}) \, $ lies entirely within the circle 
of radius
$$R^{(s)} \,\,\, = \,\,\, D \, T \, 
\sqrt{\left( m^{(s)}_{0} \right)^{2} + 
\left( n^{(s)}_{0} \right)^{2} + 
m^{(s)}_{0} n^{(s)}_{0} } $$
centered at $\, (x_{0}, y_{0}) \, $.

 It can then be seen that our component of the set 
(\ref{VminusComp}) (containing the point 
$\, (x_{0}, y_{0}) $) has the same property if
$$c \quad < \quad c_{0} \left(\alpha^{(s)}, 0, 0 \right)
\,\, - \,\, \delta_{(s)} \,\,\, , $$
where
\begin{multline}
\label{deltasVal}
\delta_{(s)} \,\,\, =  \\
= \,\,\, 4 \sqrt{3} \, C_{1} \, D \, T \,\, 
{\sqrt{\left( m^{(s)}_{0} \right)^{2} + 
\left( n^{(s)}_{0} \right)^{2} + 
m^{(s)}_{0} n^{(s)}_{0} } \over 
\left( m^{(s)} \right)^{2} } \,\,\, +  \\
+ \,\,\, {C_{1} \, T  \over 
\sqrt{3 \left( \left( m^{(s)}_{0} \right)^{2} + 
\left( n^{(s)}_{0} \right)^{2} + 
m^{(s)}_{0} n^{(s)}_{0} \right) } }
\end{multline}

 As also follows from our reasoning, for such $\, c \, $, 
the indicated property holds, in fact, for any connected 
component of the corresponding set (\ref{VminusComp}). 

 From this it can also be seen that for
$$c \quad < \quad c_{0} \left(\alpha^{(s)}, 0, 0 \right)
\,\, - \,\, \delta_{(s)} $$
the situation in the plane $\, \mathbb{R}^{2} \, $ 
for the potential $\, V (x, y, \alpha, {\bf a}) \, $ 
corresponds to the type $\, A_{-} \, $, and thus,
$$c_{0} \left(\alpha^{(s)}, 0, 0 \right)
\,\, - \,\, \delta_{(s)} \quad  <  \quad c_{0} (\alpha) $$

 Reasoning similarly in the case $\, c > c_{0} (\alpha) \, $, 
we also obtain that any connected component of the 
set (\ref{VplusComp}) has the size no greater than 
$\, D \, L^{(s)} \, $ if
$$c \quad > \quad c_{0} \left(\alpha^{(s)}, 0, 0 \right)
\,\, + \,\, \delta_{(s)} $$

 In addition, we also have the relation 
$$c_{0} \left(\alpha^{(s)}, 0, 0 \right)
\,\, + \,\, \delta_{(s)} \quad  >  \quad c_{0} (\alpha) $$

 It can be seen, therefore, that the sizes of 
the connected components (\ref{VminusComp}) and 
(\ref{VplusComp}) do not exceed $\, D \, L^{(s)} \, $, if
$$c \,\,\, \notin \,\,\, \left[ 
c_{0} (\alpha) - \Delta_{(s)} \, , \,\, 
c_{0} (\alpha) + \Delta_{(s)} \right] \,\,\, , $$
where $\,\, \Delta_{(s)} \, = \, 2 \, \delta_{(s)} \, $. 

 It is easy to see that $\, \Delta_{(s)} \rightarrow 0 \, $ 
as $\, s \rightarrow \infty \, $. Moreover, from the 
definition of the numbers 
$\, m \, $, $\, n \, $, $\, m_{0} \, $, $\, n_{0} \, $,
$\, \Delta_{(s)} \, $ and $\, L^{(s)} \, $,   
one can also see the existence of a constant $\, C \, $ 
such that for sufficiently large $\, m^{(s)} \, $ and 
$\, n^{(s)} \, $ (for all values of $\, \alpha \, $ and 
$\, {\bf a}$) we have the relation
$$D \, L^{(s)} \,\,\, \leq \,\,\, C \, \Delta_{(s)}^{-1} 
\,\,\, , $$
which implies the assertion of the lemma.

{\hfill Lemma 4.1 is proven.}

\vspace{3mm}

\noindent
{\bf Theorem 4.1}

 For almost all ``non-magic'' angles $\, \alpha \, $ 
(except for a set of measure zero) the following 
statement is true:

 For any $\, \epsilon > 0 \, $ there exists a constant 
$\, \widehat{C} \, $ such that the size of the connected 
components of the set
$$V (x, y, \alpha, {\bf a}) \,\,\, = \,\,\, c $$
does not exceed
$$\widehat{C} \, \left| c \, - \, c_{0} (\alpha) 
\right|^{- 1 - \epsilon} $$
for sufficiently small values of
$\, \left| c \, - \, c_{0} (\alpha) \right| \, $.

\vspace{2mm}

\noindent
Proof.

 Consider the above values $\, \delta_{(s)} \, $
(\ref{deltasVal}), defined by the sequence 
$\, \left( m^{(s)}, n^{(s)} \right) \, $. The numbers 
$\, \left( m^{(s)}, n^{(s)} \right) \, $ are defined 
by ``good'' approximations (\ref{wPribl}) of the 
number $\, w \, $ by the values
$\, n^{(s)} / m^{(s)} \, $, where $\, 0 < w < 1 \, $.

 According to the general theory of approximation of 
irrational numbers by rational ones 
(see, for example, \cite{Hinchin}), for almost all 
$\, w \, $ (except for a set of measure zero) 
there exist sequences $\, n^{(s)} / m^{(s)} \, $,
$$\left| {n^{(s)} \over m^{(s)}} - w \right| \,\, < \,\, 
{1 \over \left(m^{(s)}\right)^{2}} \,\,\, , $$
with denominators $\, m^{(s)} \, $ growing no faster 
(and no slower) than some geometric progressions.

 More precisely, there exist constants $\, 1 < a < A \, $ 
such that for almost all $\, w \, $ there exist sequences 
of ``good'' approximations $\, n^{(s)} / m^{(s)} \, $ 
such that
$$a \,\,\, < \,\,\, \sqrt[s]{m^{(s)}} \,\,\, < \,\,\, A $$
for sufficiently large $\, s \, $.

 Moreover, there exists a universal constant $\, \gamma \, $ 
such that
$$\sqrt[s]{m^{(s)}} \,\,\, \rightarrow \,\,\, 
\gamma \,\,\, , \quad \quad s \, \rightarrow \, \infty $$
for almost all $\, w \, $.

 For a ``typical'' angle $\, \alpha \, $ and any 
$\, \epsilon^{\prime} > 0 \, $ we can therefore put
$$\gamma \cdot \gamma^{-\epsilon^{\prime}} \quad < \quad 
\sqrt[s]{m^{(s)}} \quad < \quad 
\gamma \cdot \gamma^{\epsilon^{\prime}} \,\,\, , $$
that is
$$\gamma^{s} \cdot \gamma^{-\epsilon^{\prime} s} 
\quad < \quad  m^{(s)} \quad < \quad 
\gamma^{s} \cdot \gamma^{\epsilon^{\prime} s} $$
for sufficiently large $\, s \, $.

 Taking into account again the relations between the 
numbers $\, m^{(s)} \, $, $\, n^{(s)} \, $, 
$\, m^{(s)}_{0} \, $, $\, n^{(s)}_{0} \, $ and 
the definition of $\, \delta_{(s)} \, $ we can then 
write in this case
$$\widetilde{C}_{1} \,\, 
\gamma^{-s} \cdot \gamma^{-\epsilon^{\prime} s}
\quad < \quad \delta_{(s)} \quad < \quad 
\widetilde{C}_{2} \,\, 
\gamma^{-s} \cdot \gamma^{\epsilon^{\prime} s} $$
for some (universal for the family 
$\, V (x, y, \alpha, {\bf a}) $) constants
$\, \widetilde{C}_{1} \, $ and $\, \widetilde{C}_{2} \, $.

 Let now, for example, $\, c < c_{0} (\alpha) \, $ and
$$c \,\, \in \,\, \left[ 
c_{0} (\alpha) - \Delta_{(s)} \, , \,\,
c_{0} (\alpha) - \Delta_{(s+1)} \right] \,\, , 
\quad \Delta_{(s+1)} \, < \, \Delta_{(s)}$$
(for sufficiently large $\, s $).
 
\vspace{2mm} 
 
 According to Lemma 4.1, the sizes $\, d \, $ of 
the connected components of the set 
$\, V (x, y, \alpha, {\bf a}) \, = \, c \, $ 
do not exceed in this case
$$C \, \Delta_{(s+1)}^{-1} \,\,\, = \,\,\, 
C \, \Delta_{(s)}^{-1} \,\, 
{\delta_{(s)} \over \delta_{(s+1)}} $$

 Fixing some (small) $\, \epsilon^{\prime} > 0 \, $ 
and assuming $\, s \, $ to be large enough, 
we can then write
$$d \quad \leq \quad 
C \,\, {\widetilde{C}_{2} \over \widetilde{C}_{1}} \,\,\, 
\gamma^{1 + \epsilon^{\prime}} \, \cdot \, 
\gamma^{2 \epsilon^{\prime} s} \, \cdot \, 
\Delta_{(s)}^{-1} $$

 From the condition
$$ \delta_{(s)} \quad < \quad \widetilde{C}_{2} \,\, 
\gamma^{-s} \cdot \gamma^{\epsilon^{\prime} s} $$
we have also
$$\gamma^{(1 - \epsilon^{\prime}) s} 
\quad < \quad 2 \, \widetilde{C}_{2} \,\,
\Delta_{(s)}^{-1} \,\,\, , $$
so that
$$\gamma^{2 \epsilon^{\prime} s} \quad < \quad
\left( 2 \, \widetilde{C}_{2} \,\, \Delta_{(s)}^{-1}
\right)^{2 \epsilon^{\prime} \over 1 - \epsilon^{\prime}}
\quad = \quad \left( 2 \, \widetilde{C}_{2} 
\right)^{2 \epsilon^{\prime} \over 1 - \epsilon^{\prime}}
\Delta_{(s)}^{- {2 \epsilon^{\prime} \over 1 - \epsilon^{\prime}}}
$$

 Finally
\begin{multline*}
d \quad \leq \quad 
C \,\, {\widetilde{C}_{2} \over \widetilde{C}_{1}} \,\,\, 
\gamma^{1 + \epsilon^{\prime}} \,
\left( 2 \, \widetilde{C}_{2} 
\right)^{2 \epsilon^{\prime} \over 1 - \epsilon^{\prime}} \,\,
\Delta_{(s)}^{- 1 - {2 \epsilon^{\prime} \over 1 - \epsilon^{\prime}}}
\quad \leq  \\
\leq \quad 
C \,\, {\widetilde{C}_{2} \over \widetilde{C}_{1}} \,\,\, 
\gamma^{1 + \epsilon^{\prime}} \,
\left( 2 \, \widetilde{C}_{2} 
\right)^{2 \epsilon^{\prime} \over 1 - \epsilon^{\prime}} \,\,\,
\big| c \, - \, c_{0} (\alpha)
\big|^{- 1 - {2 \epsilon^{\prime} \over 1 - \epsilon^{\prime}}}
\end{multline*}
 (similarly for $\, c > c_{0} (\alpha)$).

\vspace{2mm}
 
 It is easy to see that for sufficiently small 
$\, \epsilon^{\prime} \, $ we can put
$$C \,\, {\widetilde{C}_{2} \over \widetilde{C}_{1}} \,\,\, 
\gamma^{1 + \epsilon^{\prime}} \,
\left( 2 \, \widetilde{C}_{2} 
\right)^{2 \epsilon^{\prime} \over 1 - \epsilon^{\prime}}
\quad \leq \quad \widehat{C} $$
for some constant $\, \widehat{C} \, $. Assuming also
$$ {2 \epsilon^{\prime} \over 1 - \epsilon^{\prime}}
\,\,\ = \,\,\, \epsilon \,\,\, , \quad \quad
\epsilon , \, \epsilon^{\prime} \, \rightarrow \, 0 
\,\,\, , $$
we obtain the statement of the theorem.

{\hfill Theorem 4.1 is proven.}

\vspace{2mm}

 In contrast to Lemma 4.1, Theorem 4.1 holds for 
``generic'' angles $\, \alpha \, $, corresponding 
to a set of full measure in the angle space. 
At the same time, there exist special $\, \alpha \, $ 
for which the numbers $\, m^{(s)} \, $ grow faster as 
$\, s \rightarrow \infty \, $, and Theorem 4.1 
is not applicable to them. For the corresponding 
potentials $\, V (x, y, \alpha, {\bf a}) \, $ 
the behavior of the function $\, d (c) \, $ as 
$\, c \rightarrow c_{0} (\alpha) \, $ can have 
a pronounced ``cascade'' structure. In particular, 
for such potentials, it may not be impossible 
to construct separate sequences 
$\, c_{(s)} \rightarrow c_{0} (\alpha) \, $ 
corresponding to the asymptotics
$$d_{(s)} \,\,\, \simeq \,\,\, 
\big| c_{(s)} \, - \, c_{0} (\alpha) \big|^{-\nu} $$
with (different) values of $\, \nu > 1 \, $ for 
$\, s \rightarrow \infty \, $. According to Lemma 4.1, 
however, such behavior is impossible in general in 
any of the intervals $\, (c , \, c_{0} (\alpha)) \, $ 
or $\, (c_{0} (\alpha) , \, c) \, $. The ``special'' 
values $\, \alpha \, $, as we have already said, 
have measure zero in the full angle space.

\vspace{2mm}

 In conclusion, we note once again that all the 
statements we made for rotational symmetry of 
the 3rd order are also transferred without 
significant changes to the case of symmetry 
of the 4th order.

\vspace{-3mm}

\section{Conclusion}
\setcounter{equation}{0}

\vspace{-3mm}

 The paper studies the behavior of level lines of 
special potentials arising in ``two-layer'' systems. 
The main subject of the study is the behavior of 
the size of closed level lines of such potentials when 
approaching the (single) level of emergence of open
level lines. It was shown that this behavior has special 
asymptotics, inherent in potentials under consideration. 
Comparison of the behavior of level lines of the 
represented potentials with the behavior of level lines 
of random potentials on a plane shows that such 
potentials can be considered as a model of random 
potentials, possessing its own specific features.

\vspace{1mm}

 The author is grateful to Prof. I.A. Dynnikov for 
fruitful discussions.


\begin{thebibliography}{99}

\bibitem{MultValAnMorseTheory} S.P. Novikov,
The Hamiltonian formalism and a many-valued analogue of  
Morse theory, {\it Russian Math. Surveys} {\bf 37} (5), 
1-56 (1982).

\vspace{1mm}

\bibitem{zorich1} A.V. Zorich,
A problem of Novikov on the semiclassical motion 
of an electron in a uniform almost rational magnetic field.,  
{\it Russian Math. Surveys} {\bf 39} (5), 287-288 (1984).

\vspace{1mm}

\bibitem{dynn1992} I.A. Dynnikov,
Proof of S.P. Novikov's conjecture for the case of 
small perturbations of rational magnetic fields,  
{\it Russian Math. Surveys} {\bf 47}:3, 172-173 (1992).

\vspace{1mm}

\bibitem{Tsarev} S.P. Tsarev, private communication, 1992-1993

\vspace{1mm}

\bibitem{dynn1} I.A. Dynnikov,
Proof of S.P. Novikov's conjecture on
the semiclassical motion of an electron,  
{\it Math. Notes} {\bf 53}:5, 495-501 (1993).

\vspace{1mm}

\bibitem{zorich2} A.V. Zorich.,
Asymptotic Flag of an Orientable Measured Foliation on a Surface.,
Proc. ``Geometric Study of Foliations''., 
(Tokyo, November 1993), ed. T.Mizutani et al. 
Singapore: World Scientific Pb. Co., 479-498 (1994).

\vspace{1mm}

\bibitem{DynnBuDA} I.A. Dynnikov.,
Surfaces in 3-torus: geometry of plane sections.,
Proc. of ECM2, BuDA, 1996.

\vspace{1mm}

\bibitem{dynn2} I.A. Dynnikov.,
Semiclassical motion of the electron. A proof of the Novikov conjecture
in general position and counterexamples., Solitons, geometry, and
topology: on the crossroad, Amer. Math. Soc. Transl. Ser. 2, 179,
Amer. Math. Soc., Providence, RI, 1997, 45-73.

\vspace{1mm}

\bibitem{dynn3} I.A. Dynnikov,
The geometry of stability regions in Novikov's problem on the
semiclassical motion of an electron,  
{\it Russian Math. Surveys} {\bf 54}:1, 21-59 (1999).

\vspace{1mm}

\bibitem{PismaZhETF} S.P. Novikov, A.Y. Maltsev,
Topological quantum characteristics observed in the 
investigation of the conductivity in normal metals,  
{\it JETP Letters} {\bf 63} (10), 855-860 (1996).

\vspace{1mm}

\bibitem{ZhETF2} A.Ya. Maltsev.,
Anomalous behavior of the electrical conductivity tensor in strong 
magnetic fields., {\it Journal of Experimental and Theoretical Physics} 
{\bf 85} (5), 934-942 (1997)

\vspace{1mm}

\bibitem{UFN} S.P. Novikov, A.Y. Maltsev,
Topological phenomena in normal metals,  
{\it Physics-Uspekhi} {\bf 41}:3, 231-239 (1998).

\vspace{1mm}

\bibitem{BullBrazMathSoc} 
A.Ya. Maltsev, S.P. Novikov.,
Quasiperiodic functions and Dynamical Systems
in Quantum Solid State Physics.,
{\it Bulletin of Braz. Math. Society}, New Series {\bf 34}:1 (2003), 
171-210.

\vspace{1mm}

\bibitem{JournStatPhys}
A.Ya. Maltsev, S.P. Novikov.,
Dynamical Systems, Topology and Conductivity in Normal Metals in
strong magnetic fields.,
{\it Journal of Statistical Physics} {\bf 115}:(1-2) (2004), 31-46.

\vspace{1mm}

\bibitem{TrMian} A.Ya. Maltsev, S.P. Novikov.,
The Theory of Closed 1-Forms, Levels of Quasiperiodic Functions 
and Transport Phenomena in Electron Systems., 
{\it Proceedings of the Steklov Institute of Mathematics}
{\bf 302}, 279-297 (2018).

\vspace{1mm}

\bibitem{NovKvazFunc} S.P. Novikov,
Levels of quasiperiodic functions on a plane, and Hamiltonian systems,  
{\it Russian Math. Surveys}, {\bf 54} (5) (1999), 1031-1032

\vspace{1mm}

\bibitem{DynNov}  I.A. Dynnikov, S.P. Novikov,
Topology of quasi-periodic functions on the plane,  
{\it Russian Math. Surveys}, {\bf 60} (1) (2005), 1-26

\vspace{1mm}

\bibitem{Zorich1996}
A.V. Zorich., 
Finite Gauss measure on the space of interval exchange transformations.
Lyapunov exponents., {\it Annales de l'Institut Fourier} {\bf 46}:2,
(1996), 325-370.

\vspace{1mm}

\bibitem{ZorichAMS1997}
Anton Zorich.,
On hyperplane sections of periodic surfaces.,
Solitons, Geometry, and Topology: On the Crossroad,
V.~M.~Buchstaber and S.~P.~Novikov (eds.),
Translations of the AMS, Ser. 2, vol. {\bf 179}, AMS, Providence, RI
(1997), 173-189.

\vspace{1mm}

\bibitem{Zorich1997}
Anton Zorich.,
Deviation for interval exchange transformations.,
{\it Ergodic Theory and Dynamical Systems} {\bf 17},
(1997), 1477-1499.

\vspace{1mm}

\bibitem{zorich3}
Anton Zorich.,
How do the leaves of closed 1-form wind around a surface.,
``Pseudoperiodic Topology'',
V.I.Arnold, M.Kontsevich, A.Zorich (eds.),
Translations of the AMS, Ser. 2, vol. 197, AMS,
Providence, RI, 1999, 135-178.

\vspace{1mm}

\bibitem{ZorichLesHouches}
Anton Zorich., 
Flat surfaces., in collect. ``Frontiers in Number
Theory, Physics and Geometry. Vol. 1: On random matrices, zeta
functions and dynamical systems''; Ecole de physique des
Houches, France, March 9-21 2003, P. Cartier; B. Julia; P.
Moussa; P. Vanhove (Editors), Springer-Verlag, Berlin, 2006,
439-586.

\vspace{1mm}

\bibitem{Skripchenko2} A. Skripchenko.,
On connectedness of chaotic sections of some 3-periodic surfaces.,
{\it Ann. Glob. Anal. Geom.} {\bf 43} (2013), 253-271.

\vspace{1mm}

\bibitem{DynnSkrip1} I. Dynnikov, A. Skripchenko.,
On typical leaves of a measured foliated 2-complex of thin type., 
Topology, Geometry, Integrable Systems, and Mathematical Physics: 
Novikov's Seminar 2012-2014, Advances in the Mathematical Sciences., 
Amer. Math. Soc. Transl. Ser. 2, 234, eds. V.M. Buchstaber, 
B.A. Dubrovin, I.M. Krichever, Amer. Math. Soc., Providence,
RI, 2014, 173-200, arXiv: 1309.4884

\vspace{1mm}

\bibitem{DynnSkrip2} I. Dynnikov, A. Skripchenko.,
Symmetric band complexes of thin type and chaotic sections which 
are not actually chaotic., {\it Trans. Moscow Math. Soc.}, Vol. 76, 
no. 2, 2015, 287-308.

\vspace{1mm}

\bibitem{AvilaHubSkrip1}
A. Avila, P. Hubert, A. Skripchenko.,
Diffusion for chaotic plane sections of 3-periodic surfaces.,
{\it Inventiones mathematicae}, October 2016, Volume 206, 
Issue 1, pp 109–146. 

\vspace{1mm}

\bibitem{DeLeo1} R. De Leo,
Existence and measure of ergodic leaves in Novikov's problem
on the semiclassical motion of an electron.,  
{\it Russian Math. Surveys} {\bf 55}:1 (2000), 166-168.

\vspace{1mm}

\bibitem{DeLeo2} R. De Leo,
Characterization of the set of ``ergodic directions'' in Novikov's
problem of quasi-electron orbits in normal metals.,  
{\it Russian Math. Surveys} {\bf 58}:5 (2003), 1042-1043.

\vspace{1mm}

\bibitem{DeLeo3} R. De Leo.,
Topology of plane sections of periodic polyhedra 
with an application to the Truncated Octahedron., 
{\it Experimental Mathematics} {\bf 15}:1 (2006), 109-124.

\vspace{1mm}

\bibitem{DeLeoDynnikov1}  R. De Leo, I.A. Dynnikov,   
An example of a fractal set of plane directions having chaotic
intersections with a fixed 3-periodic surface.,  
{\it Russian Math. Surveys} {\bf 62}:5 (2007), 990-992.

\vspace{1mm}

\bibitem{dynn4} I.A. Dynnikov,
Interval identification systems and plane sections 
of 3-periodic surfaces.,  
{\it Proceedings of the Steklov Institute of Mathematics} 
{\bf 263}:1 (2008), 65-77.

\vspace{1mm}

\bibitem{DeLeoDynnikov2} R. De Leo, I.A. Dynnikov.,
Geometry of plane sections of the infinite regular skew polyhedron
$\{ 4, \, 6 \, | \, 4 \}$., 
{\it Geom. Dedicata} {\bf 138}:1 (2009), 51-67.

\vspace{1mm}

\bibitem{Skripchenko1} A. Skripchenko.,
Symmetric interval identification systems of order three.,
{\it Discrete Contin. Dyn. Sys.} {\bf 32}:2 (2012), 643-656.

\vspace{1mm}

\bibitem{AvilaHubSkrip2}
A. Avila, P. Hubert, A. Skripchenko.,
On the Hausdorff dimension of the Rauzy gasket.,
{\it Bulletin de la societe mathematique de France}, 
2016, {\bf 144} (3), pp. 539 - 568.

\vspace{1mm}

\bibitem{AnnPhys}
A.Ya. Maltsev, S.P. Novikov, 
Open level lines of a superposition of periodic potentials on a plane, 
{\it Annals of Physics} {\bf 447}(Pt.2), 169039 (2022)

\vspace{1mm}

\bibitem{DynHubSkrip}
Ivan Dynnikov, Pascal Hubert, Alexandra Skripchenko,
Dynamical Systems Around the Rauzy Gasket and Their Ergodic Properties,
{\it International Mathematics Research Notices IMRN}  
Int. Math. Res. Not. Volume 2023, Issue 8, 6461–6503 (2023); 
arXiv 2011.15043

\vspace{1mm}

\bibitem{Shallcross1}
S. Shallcross, S. Sharma, and O. A. Pankratov,
Quantum Interference at the Twist Boundary in Graphene,
{\it Phys. Rev. Lett} {\bf 101}, 056803 (2008)

\vspace{1mm}

\bibitem{Shallcross2}
S. Shallcross, S. Sharma, E. Kandelaki, and O. A. Pankratov,
Electronic structure of turbostratic graphene,
{\it Phys. Rev. B}  {\bf 81}, 1 (2010)

\vspace{1mm}

\bibitem{GeimGrigorieva}
A.K. Geim, I.V. Grigorieva,
Van der Waals heterostructures,
{\it Nature} {\bf 499}, 419–425 (2013)

\vspace{1mm}

\bibitem{RShRN}
A.V. Rozhkov, A.O. Sboychakov, A.L. Rakhmanov, Franco Nori,
Electronic properties of graphene-based bilayer systems,
{\it Physics Reports} {\bf 648}, 1-104 (2016)

\vspace{1mm}

\bibitem{CMFCLK}
Stephen Carr, Daniel Massatt, Shiang Fang, Paul Cazeaux,
Mitchell Luskin, and Efthimios Kaxiras,
Twistronics: Manipulating the electronic properties of
two-dimensional layered structures through their twist angle,
{\it Phys. Rev. B} {\bf 95}, 075420 (2017)

\vspace{1mm}

\bibitem{KhalafKruchTarnVish}
Eslam Khalaf, Alex J. Kruchkov, Grigory Tarnopolsky, 
and Ashvin Vishwanath,
Magic Angle Hierarchy in Twisted Graphene Multilayers,
{\it Phys. Rev. B} {\bf 100}, 085109 (2019)

\vspace{1mm}

\bibitem{DindorkarKuradeShaikh}
Shreyas S. Dindorkar, Ajinkya S. Kurade, Aksh Hina Shaikh,
Magical moir\'e patterns in twisted bilayer graphene: 
A review on recent advances in graphene twistronics,
{\it Chemical Physics Impact} {\bf 7}, 100325 (2023)

\vspace{1mm}

\bibitem{PaulCrowleyFu}
Nisarga Paul, Philip J. D. Crowley, and Liang Fu, 
Directional Localization from a Magnetic Field in Moir\'e
Systems, {\it Phys. Rev. Lett.} {\bf 132}, 246402 (2024)

\vspace{1mm}

\bibitem{BernevigEfetov}
B. Andrei Bernevig, Dmitri K. Efetov,
Twisted bilayer graphene’s gallery of phases,
{\it Physics Today} {\bf 77} (4), 38–44 (2024)

\vspace{1mm}

\bibitem{TitovKatsnelson}
M. Titov and M.I. Katsnelson,
Metal-Insulator Transition in Graphene on Boron Nitride,
{\it Phys. Rev. Lett.} {\bf 113}, 096801 (2014).

\vspace{1mm}

\bibitem{Superpos} A.Ya. Maltsev, 
On the Novikov problem for superposition of periodic potentials, 
arXiv:2409.09759 

\vspace{1mm}

\bibitem{DynMalNovUMN} I.A. Dynnikov, A.Ya. Maltsev, S.P. Novikov,
Geometry of quasi-periodic functions on the plane,  
{\it Russian Math. Surveys} {\bf 77} : 6, 1061–1085
(2022), arXiv:2306.11257

\vspace{1mm}

\bibitem{BigQuas} A.Ya. Maltsev, 
On the Novikov problem with a large number of quasiperiods 
and its generalizations, 
Proceedings of the Steklov Institute of Mathematics, 
2024, Vol. 325, pp. 163–176, 
arXiv:2309.01475

\vspace{1mm}

\bibitem{Stauffer} 
D. Stauffer, Scaling theory of percolation clusters,
{\it Physics Reports}, Volume 54, Issue 1 (1979), 1-74.

\vspace{1mm}

\bibitem{Essam} 
J.W. Essam, Percolation theory,
{\it Rep. Prog. Phys.} {\bf 43} (1980), 833-912

\vspace{1mm}

\bibitem{Riedel} Eberhard K. Riedel, 
The potts and cubic models in two dimensions: 
A renormalization-group description,
{\it Physica A: Statistical Mechanics and its Applications.,}
Volume 106, Issues 1-2 (1981), 110-121.

\vspace{1mm}

\bibitem{Trugman} S.A. Trugman,
Localization, percolation, and the quantum Hall effect,
{\it Phys. Rev. B} {\bf 27} (1983), 7539-7546

\vspace{1mm}

\bibitem{WZLWLMHXMCJ} 
Yunfei Wang, Jia-Hui Zhang, Yuqing Li, Jizhou Wu, Wenliang Liu, 
Feng Mei, Ying Hu, Liantuan Xiao, Jie Ma, Cheng Chin, and Suotang Jia,
Observation of Interaction-Induced Mobility Edge in an Atomic Aubry-Andre 
Wire, {\it Phys. Rev. Lett.} {\bf 129}, 103401 (2022).

\vspace{1mm}

\bibitem{LesserLifshitz}
Omri Lesser and Ron Lifshitz,
Emergence of quasiperiodic Bloch wave functions in quasicrystals,  
{\it Physical Review Research} {\bf 4} (1), 013226 (2022). 

\vspace{1mm}

\bibitem{Hinchin} A.Ya. Hinchin,
Continued fractions., Moscow: FISMATGIS, (1961).

\end{thebibliography}
\end{document}